\documentclass[twocolumn,preprintnumbers]{revtex4}

\usepackage[dvipdfmx]{hyperref}
\usepackage{graphicx}
\usepackage{epsf}
\usepackage{amsmath,amssymb}
\usepackage{color}

\def\Pb#1#2#3{{}^{#1#2#3}\textrm{Pb}}
\def\Ni#1#2{{}^{#1#2}\textrm{Ni}}
\def\Ca#1#2{{}^{#1#2}\textrm{Ca}}
\def\Sn#1#2#3{{}^{#1#2#3}\textrm{Sn}}

\newcommand{\PbA}{{}^{A}\textrm{Pb}}
\newcommand{\nhole}{h}

\usepackage{ulem}

\begin{document}

\preprint{KUNS-2876}
\title{Isotopic analysis of 295 MeV proton scattering off $^{204,206,208}$Pb for 
improvement of neutron densities and radii}

\author{Yoshiko Kanada-En'yo}
\affiliation{Department of Physics, Kyoto University, Kyoto 606-8502, Japan}

\begin{abstract}
A new method of reaction analysis for proton elastic scattering is proposed, 
combining systematic analyses of nuclear structure and reactions in a series of isotopes. 
This method is applied to $\Pb204(p,p)$, $\Pb206(p,p)$, and $\Pb208(p,p)$ at $E_p=295$~MeV to obtain
improved neutron densities and radii from experimental cross section data.
The reaction calculation is performed according to the relativistic impulse approximation using the 
modified Murdock and Horowitz model with  density-dependent effective $NN$ interaction.
Analysis of Pb isotopes is performed using 
theoretical densities given by the relativistic Hartree-Bogoliubov and nonrelativistic Skyrme Hartree-Fock-Bogoliubov
calculations and the experimental density extracted from the 
$(p,p)$ data at 295~MeV. 
The isotopic ratios ($R(\sigma)$) of the $\PbA(p,p)$-to-$\Pb208(p,p)$ cross sections
are analyzed in connection with the 
isotopic differences ($D(\rho_n)$ and $D(r_n)$) in neutron densities $(\rho_n(r))$ and radii $(r_n)$
between $\PbA$ and $\Pb208$.
A hole-model analysis is performed,  
assuming $\Pb204$ and $\Pb206$ as a $\Pb208$ core and neutron holes
to clarify the one-to-one correspondence between  isotopic structure properties such as $D(\rho_n)$ and $D(r_n)$ 
and the isotopic cross-section ratio. 
By fitting $R(\sigma)$, the values of $D(\rho_n)$ and $D(r_n)$ are
extracted from the experimental $(p,p)$ cross section data with less uncertainty of the structure and reaction models. 
The isotopic neutron radius difference between $\Pb206$ and $\Pb208$ was obtained as 
$D(r_n)=-0.012$~fm with an acceptable range of $D(r_n)=-0.03 \sim -0.006$~fm. 
\end{abstract}

\maketitle

\section{Introduction}
The neutron skin thickness $\Delta r_{np}$ in $N>Z$ nuclei---which is defined as the difference between the
root-mean-square~(rms) point-proton ($r_p$) and neutron $(r_n)$ radii $\Delta r_{np}=r_n-r_p$---
has long been intensively
investigated through experimental and theoretical studies.
The experimental determination of $\Delta r_{np}$ in double closed nuclei such as $\Ca48$, $\Pb208$, and $\Sn132$
is an urgent issue for accessing to the neutron matter EOS via the connection between 
$\Delta r_{np}$ and the symmetry energy parameters~\cite{RocaMaza:2011pm,Roca-Maza:2018ujj,Tsang:2012se}. 
Various experimental approaches have been applied for measuring $\Delta r_{np}$. 
For stable nuclei, point-proton radii and density distributions can be precisely determined 
from the charge radii and form factors measured by electric probes such as the isotope shift of $x$ rays and electron scattering.
However, measuring neutron radii is relatively difficult; hence,
the $\Delta r_{np}$ data still contain large experimental errors.  
Many experimental efforts have been made to determine the $r_n$ of $\Pb208$ by measuring 
proton elastic scattering \cite{Starodubsky:1994xt,Zenihiro:2010zz}, 
$x$-ray cascade from antiprotonic atoms \cite{Klos:2007is}, parity-violating electron
scattering \cite{Abrahamyan:2012gp}, and pionic probes~\cite{Friedman:2012pa,Tarbert:2013jze}.
The electric dipole polarization measured by 
polarized proton inelastic scattering \cite{Tamii:2011pv} is an alternative approach for determining 
$\Delta r_{np}$~\cite{Piekarewicz:2012pp}.

The observed $\Delta r_{np}$ data were used to extract information for the neutron matter EOS based on 
nuclear density functional theory~(DFT) by comparing the data with theoretical predictions of $\Delta r_{np}$
calculated using a variety of relativistic and nonrelativistic energy density functionals. 
In extracting infinite matter properties from nuclear observables using DFT, it is essential to reduce model dependences 
mainly originating in finite-size effects 
such as surface contributions and shell effects in $\Delta r_{np}$~\cite{Centelles:2010qh}.

Proton elastic scattering is a useful tool for investigating the detailed profile of the neutron-density distribution
as used for nuclei in wide mass-number regions.
The neutron-density distribution of $\Pb208$ was extracted from the $(p,p)$ data 
at 800 MeV~\cite{Ray:1978ws,Ray:1979qv,Hoffmann:1980kg}, 650~MeV~\cite{Starodubsky:1994xt}, and 295~MeV~\cite{Zenihiro:2010zz}. 
For reaction analyses of proton elastic scattering at several hundreds of MeV, 
a reaction model based on the relativistic impulse approximation (RIA) framework with 
meson-exchange effective $NN$ interactions was proposed by Murdock and Horowitz (MH model); 
this model is based on globally fitting the $(p,p)$ data 
in the $E_p=100$--400~MeV energy range~\cite{Horowitz:1985tw,Murdock:1986fs,RIAcode:1991}.
The RIA code with the MH model has been widely used to investigate neutron density based on $(p,p)$ data.
Later, an improved version, the density-dependent MH model (ddMH) was proposed by Sakaguchi and his collaborators,
introducing density dependences of the effective $NN$ interaction~\cite{Sakaguchi:1998zz,Terashima:2008zza,Zenihiro:2010zz}. 
The ddMH model was finely calibrated to 295 MeV proton elastic scattering off $\Ni58$ 
at scattering angles of $\theta_\textrm{c.m.}$ 
up to $\sim 50^\circ$, and 
 proton elastic scattering off 
the Sn~\cite{Terashima:2008zza}, Pb~\cite{Zenihiro:2010zz}, and Ca~\cite{Zenihiro:2018rmz} isotopes was 
successfully described.
By taking high-quality measurements of the proton elastic scattering off $\Pb204$, $\Pb206$, and $\Pb208$ at 295~MeV, 
the neutron densities were extracted via reaction analysis using RIA with the ddMH model and  
the $r_n$ values in a series of Pb isotopes were obtained~\cite{Zenihiro:2010zz}. 

However, the $r_n$ values of the Pb isotopes determined by the $(p,p)$ data 
still contain significant experimental errors, mainly because of the large uncertainty of the extracted neutron density in the internal region, 
to which proton scattering at several hundreds of MeV is insensitive~\cite{Piekarewicz:2005iu}. 
To avoid the uncertainty in $r_n$ extracted from the $(p,p)$ data,  
a new method based DFT was proposed to evaluate the symmetry-energy parameters by 
comparing the surface-neutron-density profile of the theoretical predictions with 
the experimental density extracted from the data instead of $r_n$,
and it was applied to analysis of the $\Pb208$ and $\Ca48$ densities \cite{PhysRevC.102.064307}.
Alternatively, a more direct approach between the data and matter information
through a comparison of the experimental $(p.p)$ cross sections with theoretical predictions 
obtained by reaction calculations using DFT densities may be worth considering.

Our aim in this paper is to propose a new reaction analysis method for the proton elastic scattering 
measured in a series of isotopes to obtain improved neutron density and radii from the data.
Because of the similarities between neighboring nuclei, a perturbative treatment is possible, and 
relative differences between neighboring nuclei can be easily detected with less uncertainty in principle.   
Therefore, a systematic analysis combining the data in a series of neighboring isotopes can be a better tool 
than independent analyses because it can significantly reduce systematic errors.
A similar concept is often used, for instance, to determine of charge-radius differences 
by measuring the isotope shift of $x$ rays.
For proton scattering, an isotopic analysis was performed to discuss neutron-density differences~\cite{Ray:1979qv};
however, an important innovation of the present work  is that it combines the isotopic properties 
of nuclear structure with those of nuclear reaction---that is an isotopic analysis of the scattering cross sections is performed to extract structural differences 
between isotopes via reaction calculations. 
For electron elastic scattering, an isotopic analysis of the cross sections
has been performed to discuss the isotopic difference in the charge radius~\cite{RocaMaza:2008cg};
however, its application is limited. 
Little work has been done on proton elastic scattering, except for a brief 
discussion of the isotopic cross-section ratio and size scaling of the proton-nucleus potential~\cite{Hoffmann:1990ve}.

In this paper, I propose a new approach using isotopic analysis of the proton elastic scattering cross sections.
Based on isotopic systematics in nuclear structures and reactions. 
In the ground state of even-even nuclei, 
the nuclear structure changes smoothly in a series of isotopes without drastic changes, 
other than crossing a shell gap or phase transition for nuclear deformation.
Moreover, the reaction processes of neighboring isotopes at the same energy 
should be similar, provided that the projectile energy is high enough to 
neglect the mass-number dependence of higher-order effects such as channel-coupling effects
in the proton scattering.
Another advantage of this approach is that, experimental systematic errors can be significantly reduced
using the observed data in a series of isotopes measured  experimentally with  
the same setup in the same facility.

For isotopic analysis, 
the reaction calculations of $\Pb204(p,p)$, $\Pb206(p,p)$, and $\Pb208(p,p)$ at $E_p=295$~MeV
were performed via RIA calculation using the ddMH model in the same way as was
done in the analysis of Ref.~\cite{Zenihiro:2010zz}. 
As for the target densities in the reaction calculation, 
the theoretical densities of the Pb isotopes obtained by
relativistic Hartree-Bogoliubov~(RHB) and nonrelativistic Skyrme Hartree-Fock-Bogoliubov~(SHFB)
calculations of spherical nuclei
and the experimental density extracted by fitting 
the $(p,p)$ data are used.
The isotopic neutron density and radius differences and the isotopic cross-section ratios 
of $\Pb204$, $\Pb206$, and $\Pb208$ are investigated. 
A detailed analysis is performed by introducing a hole model of 
of $\Pb204$ and $\Pb206$ as a  $\Pb208$ core and neutron holes, 
the sensitivity of the cross sections to surface-neutron densities is clarified. 
Through the hole-model analysis, 
the improved neutron densities and radii are obtained by fitting the isotopic cross-section ratio obtained from the $(p,p)$ data.
The uncertainty in the structure and reaction models is discussed. 

This paper is organized as follows. The structure and reaction calculations 
are described and the obtained results of the Pb isotopes are presented in Sec.~\ref{sec:calculation}.
Isotopic analysis using the theoretical and experimental densities is performed in Sec.~\ref{sec:analysis1}, 
and the hole-model analysis is done in Sec.~\ref{sec:analysis2}.
In Sec.~\ref{sec:ndens}, improved neutron densities and radii are presented.  
A summary is given in Sec.~\ref{sec:summary}. 
In Appendix \ref{sec:app1}, 
supplemental figures are presented.

\section{Calculation of $p$ scattering} \label{sec:calculation}
\subsection{Target densities}
The theoretical densities of Pb isotopes, which are used as inputs for nucleon-nucleus folding potentials in Pb$(p,p)$ calculations, 
are obtained via the RHB and SHFB calculations of spherical nuclei using 
the computational codes named DIRHB~\cite{Niksic:2014dra} and 
HFBRAD~\cite{Bennaceur:2005mx}, respectively.
For the former, the 
DD-ME2~\cite{Lalazissis:2005de} and DD-PC1~\cite{Niksic:2008vp} interactions are used; in this paper, 
these are simply denoted by me2 and pc1, respectively. 
For the latter case, the SKM*~\cite{Bartel:1982ed} and SLy4~\cite{Chabanat:1997un} interactions are used.
Note that the parameter sets of these structure models were adjusted to
globally fit the binding energies and rms charge radii in wide 
mass-number regions that include $\Ca40$ and $\Pb208$. 

The experimental neutron densities of the Pb isotopes used in this paper are those of Refs.~\cite{Zenihiro:2010zz,Zenihiro:dron},
which were extracted by fitting the Pb$(p,p)$ data at 295 MeV using the RIA calculation 
with the ddMH model, which was  called medium-modified RIA calculation in the original paper. 
The neutron-density distribution is written in a sum-of-Gaussians (SOG) form and called the SOG-fit density in the present paper. 
The experimental proton densities are taken from 
Ref.~\cite{Zenihiro:2010zz}; they were obtained by unfolding the nuclear charge distribution determined by 
electric elastic scattering data \cite{DeJager:1987qc}.

\begin{figure}[!h]
\includegraphics[width=0.5\textwidth]{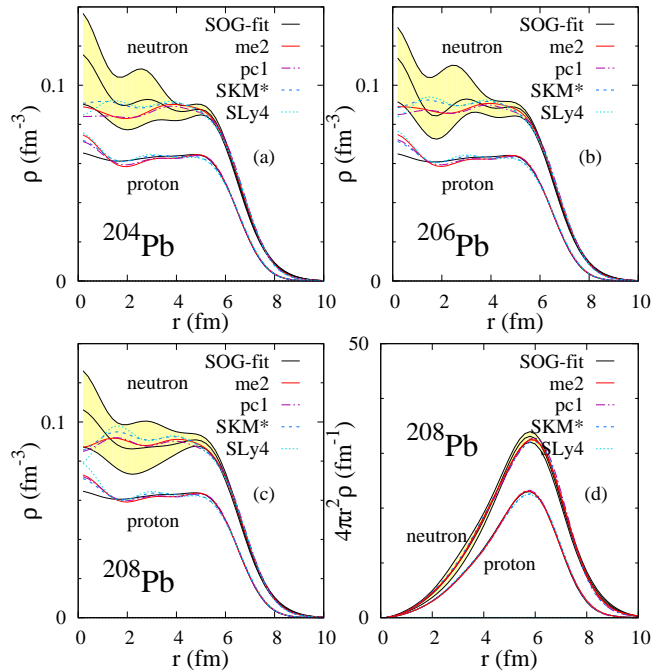}
  \caption{Proton and neutron densities $\rho_p(r)$ and $\rho_n(r)$ for  (a)~$\Pb204$, (b)~$\Pb206$,  (c)~$\Pb208$,
and (d)~$4\pi r^2 \rho_n(r)$ of $\Pb208$. 
Theoretical densities are those of RHB (me2 and pc1) and SHFB (SKM* and SLy4) 
calculations.
The experimental SOG-fit densities are the neutron density (with error envelopes) 
extracted from the $(p,p)$ data at $E_p=295$~MeV in Ref.~\cite{Zenihiro:dron}
and the proton density of Ref.~\cite{Zenihiro:2010zz} 
obtained by unfolding the charge densities determined by electron scattering data \cite{DeJager:1987qc}.
In practice, the  $\rho_p(\Pb208;r)$ data are read from figures of Ref.~\cite{Zenihiro:2010zz}, 
whereas $\rho_n(\Pb204;r)$ and $\rho_p(\Pb206;r)$ are 
constructed by $r$ scaling of $\rho_p(\Pb208;r)$ to 
adjust the point-proton radii.
\label{fig:dens-pn-1}}
\end{figure}

Figures \ref{fig:dens-pn-1}(a)-(c) show proton ($\rho_p$) and neutron ($\rho_n$) densities of $\Pb204$, $\Pb206$, and 
$\Pb208$, and Fig.~\ref{fig:dens-pn-1}(d) shows $4\pi r^2\rho_p$ and $4\pi r^2\rho_n$ of $\Pb208$.
The theoretical densities of the RHB (me2 and pc1) and SHFB (SKM* and SLy4) calculations and 
the experimental SOG-fit density are shown. 
The proton scattering at $E_p=295$~MeV is a sensitive probe of  the surface-neutron density around the peak position of 
$4\pi r^2\rho_n$ at $r\sim 6$ fm, but it is insensitive to the inner densities. 
Therefore, the SOG-fit neutron densities of Pb isotopes have large uncertainties in the $r <4$ fm region
as shown by the error envelopes (filled area).
Theoretical neutron densities depend upon structure models, but each model yields
similar neutron densities between $\Pb204$, $\Pb206$, and $\Pb208$.
Comparing the theoretical model and SOG-fit neutron densities in the surface region, 
the me2 model shows the best agreement with the SOG-fit density, whereas other theoretical models 
show disagreements;  
$4\pi r^2\rho_n(r)$ obtained by the pc1, SKM*, and SLy4 models exhibits a peak position that is slightly shifted outward compared with 
the SOG-fit and me2 densities. This peak-position shift causes deviation of the $(p,p)$ cross sections from the data 
at backward angles, as discussed later in Sec.~\ref{subsec:reaction}.

\begin{figure}[!htp]
\includegraphics[width=7cm]{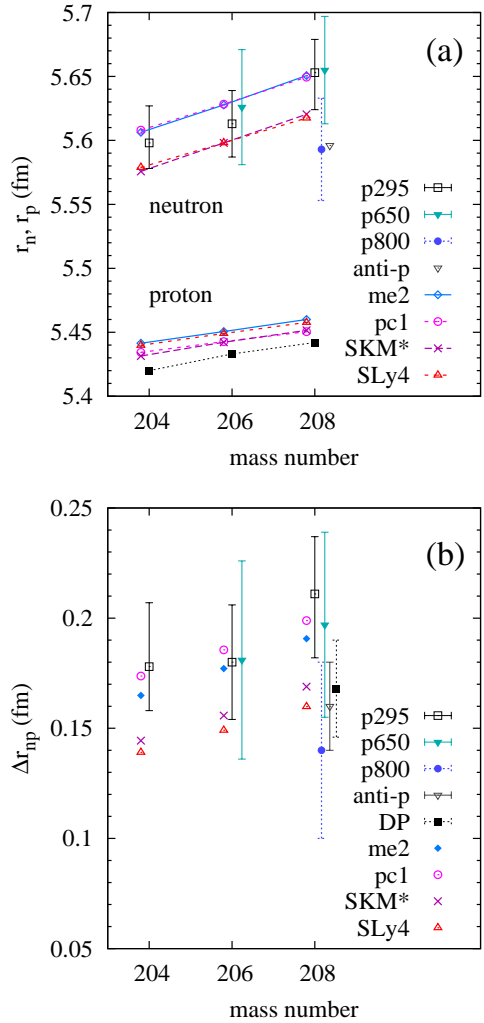}
  \caption{(a) Rms point-proton and neutron radii and (b) neutron skin thickness of Pb isotopes.
The theoretical values are those of RHB (me2 and pc1) and SHFB (SKM* and SLy4) 
calculations.
The experimental data are values of the SOG-fit density extracted from the $(p,p)$ data at 295 MeV~\cite{Zenihiro:2010zz}, 
those extracted from the
$(p,p)$ data at 800 MeV~\cite{Ray:1978ws,Ray:1979qv,Hoffmann:1980kg} and 650 MeV \cite{Starodubsky:1994xt}, and 
the $x$-ray cascade from antiprotonic atoms\cite{Klos:2007is}. The experimental $\Delta r_{np}$ value extracted 
from electric dipole polarization~\cite{Tamii:2011pv,Piekarewicz:2012pp} is also shown in (b).
\label{fig:rmsr}}
\end{figure}

Figure \ref{fig:rmsr}(a) shows the theoretical and experimental values of $r_p$ and $r_n$, whereas 
Fig.~\ref{fig:rmsr}(b) shows the values of $\Delta r_{np}$. 
All theoretical models show smooth changes of $r_p$ and $r_n$ as the mass
number $A$ increases from $\Pb204$ to $\Pb208$. The $A$ dependence of $r_p$ is quite similar between different models and 
is consistent with the experimental data. 
As for the neutron radii, the $r_n$ and $\Delta r_{np}$ change smoothly with the same slope as $A$ increases
in all theoretical models, meaning that the $A$ dependence is approximately model independent though 
the absolute $r_n$ values are model dependent. 
Compared with the smooth changes of the theoretical $r_n$ and $\Delta r_{np}$ values, 
the $r_n$ values of the SOG-fit density seem to show a different $A$ dependence---in particular, 
an enhancement of $r_n$ and $\Delta r_{np}$ at $A=208$ from $\Pb206$ to $\Pb208$. However, 
this is not definite due to the large experimental errors that arise mainly from 
uncertainty in the internal neutron densities extracted from the $(p,p)$ data.

\subsection{Reaction calculation of $p$ scattering at 295 MeV} \label{subsec:reaction}

Proton elastic scattering at $E_p=295$~MeV is calculated using RIA with the ddMH model
(a modified MH model proposed by Sakaguchi {\it et al.} \cite{Sakaguchi:1998zz} 
by introducing density-dependent $\sigma$- and $\omega$-meson masses and coupling constants 
of the relativistic Love-Franey (RLF) $NN$ interaction of the original MH model~\cite{Horowitz:1985tw,Murdock:1986fs}).
The density dependence is considered as ``medium effects'' of the effective $NN$ interaction,
which includes various many-body effects in proton elastic scattering such as 
the Pauli blocking and multistep processes, in addition to the medium effects of meson properties. 
The RIA calculation with the ddMH model is performed in the default case, and 
that with the MH model is also performed in an optional case.

The parameter set of the density dependence used in this paper is 
the same as that used to analyze Pb$(p,p)$ and Ca$(p,p)$ at $E_p=295~$MeV in Refs.~\cite{Zenihiro:2010zz,Zenihiro:2018rmz}.
This parameter set is the latest version calibrated 
to fit the updated data for $p+^{58}\textrm{Ni}$ at $E_p=295~$MeV 
and can well reproduce the $\Ca40(p,p)$ data at $E_p=295~$MeV~\cite{Zenihiro:2018rmz}.
In the RIA framework with the MH and ddMH models, the proton-nucleus potentials are calculated by folding the vector 
and scalar densities of 
target nuclei with the meson-exchange $NN$ interaction. As the input target densities, the proton $(\rho_p(r))$ and neutron 
$(\rho_n(r))$ densities are used for the proton and neutron vector densities, whereas $0.96\rho_p(r)$ 
and  $0.96\rho_n(r)$ are used for the proton and neutron scalar densities, respectively;
this prescription of the scalar densities was adopted in Ref.~\cite{Zenihiro:2010zz} to 
fit the ddMH model to the $\Ni58(p,p)$ data and applied to the Pb$(p,p)$ analysis. 
Note that, in the RHB calculations, the scalar densities can be obtained without such an approximation,
but  give only a minor correction to the Pb$(p,p)$ reactions at this energy. 
We adopt the prescription of the scalar densities for all models
consistently with the calibration of the ddMH model. 

\begin{figure}[!h]
\includegraphics[width=8 cm]{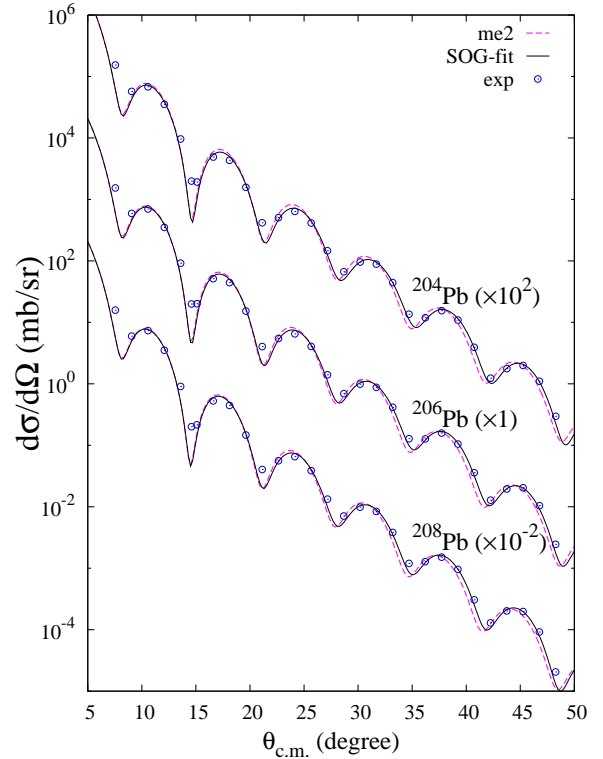}
  \caption{Cross sections of $\textrm{Pb}(p,p)$ at 295 MeV obtained by RIA with the ddMH model 
using the SOG-fit and me2 densities together with the experimental data~\cite{Zenihiro:2010zz}. 
\label{fig:cross}}
\end{figure}

\begin{figure}[!h]
\includegraphics[width=0.5\textwidth]{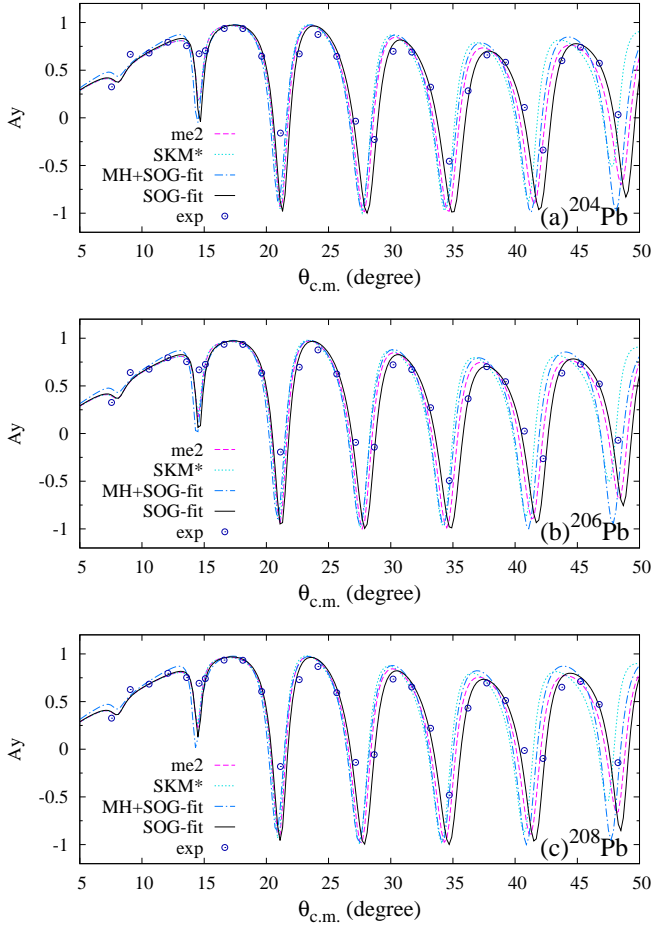}
  \caption{Analyzing powers of Pb$(p,p)$ at 295 MeV obtained by RIA with the ddMH model using the SOG-fit, me2, and SKM* densities together with the experimental data from~\cite{Zenihiro:2010zz}.
The result of the original MH model using the SOG-fit density is also 
shown for comparison.  
\label{fig:Ay}}
\end{figure}

\begin{figure}[!h]
\includegraphics[width=0.5\textwidth]{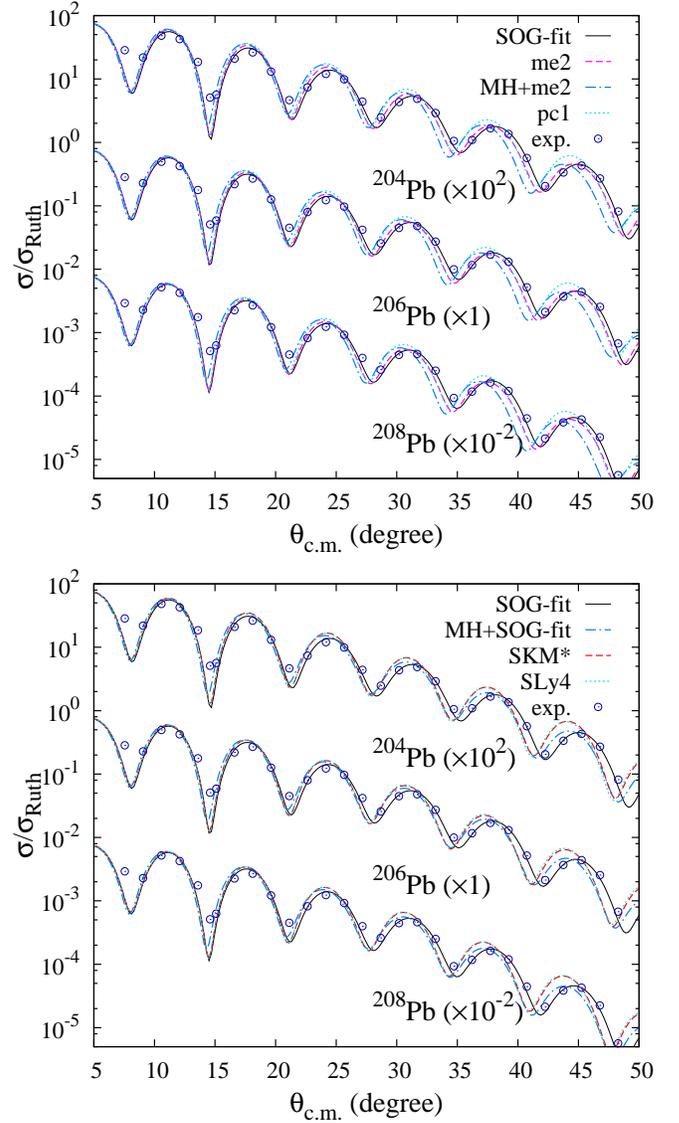}
  \caption{Rutherford ratios of the Pb$(p,p)$ cross sections at 295 MeV obtained with the ddMH model 
using the SOG-fit, me2, pc1, SKM, and SLy4 densities together with the experimental data 
from Ref.~\cite{Zenihiro:2010zz}. 
The results obtained using the original MH model using the SOG-fit (MH+SOG-fit) and me2 (MH+me2) densities 
are also shown for the purposes of comparison.
\label{fig:cross-ratio-pc1}}
\end{figure}

The cross sections and analyzing powers of Pb$(p,p)$ at 295~MeV obtained by the RIA calculation with the ddMH model
are shown in Figs.~\ref{fig:cross} and \ref{fig:Ay}, respectively, and the Rutherford ratio of the cross sections are
shown in Fig.~\ref{fig:cross-ratio-pc1}.  
In the results obtained using the pc1, SKM*, and SLy4 densities, 
dip and peak positions in the diffraction 
pattern deviate from the experimental data, in particular, at backward angles, 
in which  the dip (peak) positions shift to forward angles 
because of the slight outward shift of the $4\pi r^2\rho_n(r)$ peak position 
compared with the SOG-fit and me2 densities.
The me2 result agrees better with the data than those of other theoretical densities, 
but a slight deviation from the data still remains. Therefore, 
the dip interval in the diffraction pattern is sensitive to the $4\pi r^2\rho_n(r)$ peak position, but 
does not necessarily correspond to the neutron radius $r_n$. 
For instance, the SKM* density has a smaller value of $r_n$ than the SOG-fit density, but 
it gives a shrunk diffraction pattern of the $(p,p)$ cross sections 
consistent with the outward shift of the $4\pi r^2\rho_n(r)$ peak, indicating an expansion of  
the nuclear size probed by proton scattering. 

To show reaction model dependence, 
the results obtained using the MH model 
with the SOG-fit and me2 densities are shown in Figs.~\ref{fig:Ay} and \ref{fig:cross-ratio-pc1}
in comparison with the results of the ddMH model.
The dip interval of the cross sections depends upon the effective $NN$ interaction in the 
reaction model; compared with the 
ddMH results, the diffraction pattern of the MH model deviates forward, indicating that the range 
of the effective $NN$ interaction in the MH model is slightly longer than that in the ddMH model.

\begin{figure}[!h]
\includegraphics[width=0.5\textwidth]{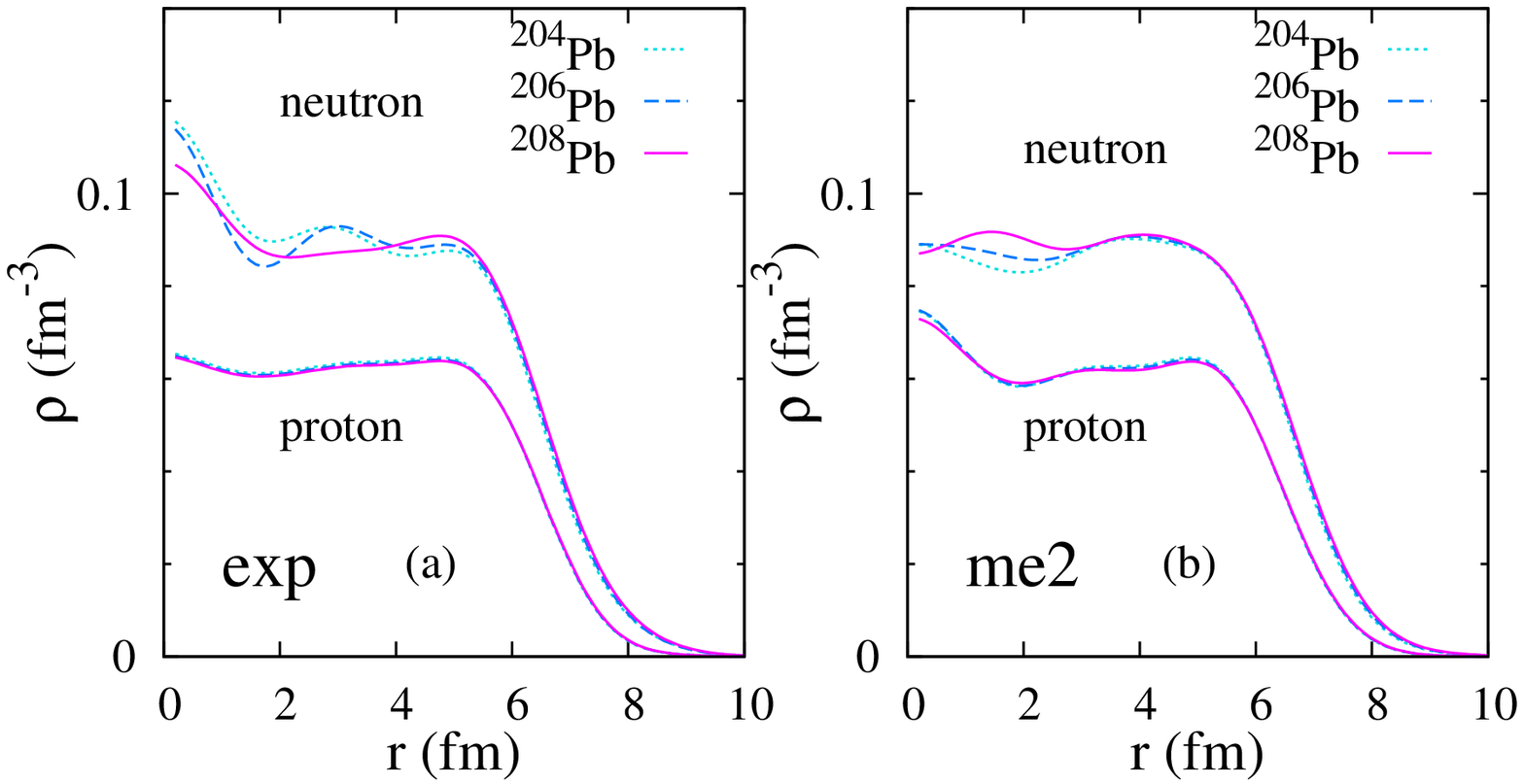}
  \caption{(a) Experimental SOG-fit densities and (b) theoretical (me2) densities of the  protons and neutrons of Pb isotopes.
\label{fig:dens-pn-2}}
\end{figure}

\begin{figure}[!h]
\includegraphics[width=0.5\textwidth]{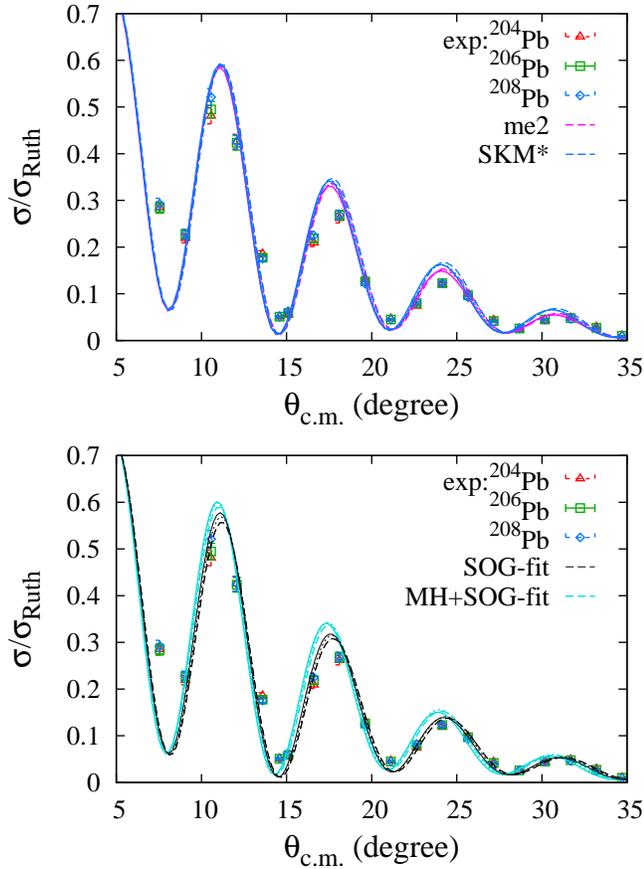}
  \caption{Rutherford ratio of the $(p,p)$ cross sections at 295 MeV, as obtained by RIA with the ddMH model
using the SOG-fit, me2, and SKM* densities together with the experimental data~\cite{Zenihiro:2010zz}.
The results obtained using the original MH model using the SOG-fit (MH+SOG-fit) and me2 (MH+me2) densities 
are also shown. The calculated Rutherford ratios of the $\Pb204(p,p)$, 
$\Pb206(p,p)$, and $\Pb208(p,p)$ cross sections are plotted using dashed, dotted, and solid lines, respectively. 
\label{fig:cross-ratio-linear}}
\end{figure}

As shown in the results, the calculated $(p,p)$ cross sections at 295~MeV are sensitive to  
differences between input neutron densities at the nuclear surface, 
and depend upon the effective $NN$ interaction used in the nucleon-nucleus folding potential. However, 
in each model, 
the results for a series of isotopes from $\Pb204$ to $\Pb208$ 
are quite similar and 
deviation from the experimental data occurs systematically for the three isotopes.
To see the isotopic similarities, 
the results of Pb isotopes are compared in Fig.~\ref{fig:dens-pn-2} for densities and Fig.~\ref{fig:cross-ratio-linear}
for the Rutherford ratio of the cross sections on a linear scale. 
For each result of the SOG-fit, me2, pc1, SKM*, SLy4, MH-SOG-fit, and MH-me2, 
the isotopic difference between $\Pb204$, $\Pb206$, and $\Pb208$ in the cross sections is 
small because surface densities in the $r\gtrsim 6$ fm region of the three isotopes approximately coincide,. 
even in the linear plot of the Rutherford ratio (Fig.~\ref{fig:cross-ratio-linear}).
These isotopic systematics are useful for a model-independent analysis of proton elastic scattering.

\clearpage

\section{Isotopic analysis} \label{sec:analysis1}

\subsection{Definitions of isotopic differences and ratios}
By considering $\Pb208$ as the reference nucleus, the isotopic density and radius differences 
and cross-section ratio are calculated for each model in the present isotopic analysis. 
The isotopic density and radius differences are respectively given as 
\begin{align}
&D(\rho_{n,p};r)\equiv \rho_{n,p}(^A\textrm{Pb};r)-\rho_{n,p}(\Pb208;r),\\
&D(r_{n,p})\equiv r_{n,p}(^A\textrm{Pb})- r_{n,p}(\Pb208),\nonumber \\
&= \frac{\int 4\pi r^4 D(\rho_{n,p};r) dr}{A-208} \label{eq:dr-drho}
\end{align}
where $\rho_{n(p)}(\PbA^A;r)$ and $r_{n(p)}(^A\textrm{Pb})$ are the neutron(proton) density
and rms radius of $^A\textrm{Pb}$. 
The isotopic cross-section ratio is given as 
\begin{align}
& R(\sigma;\theta_\textrm{c.m.})\equiv \frac{d\sigma(\PbA)/d\Omega}{d\sigma(\Pb208)/d\Omega},
\end{align}
with differential cross sections in the center-of-mass frame.
It is trivial that, even in the case of $\rho_{p,n}(^A\textrm{Pb};r)\propto \rho_{p,n}(\Pb208;r)$, 
$R(\sigma;\theta)$ is not constant 
because of normalization of the target densities used for the folding potential and kinematical effects in the reaction.

Experimental values of $R(\sigma;\theta_\textrm{c.m.})$ are obtained 
using the $(p,p)$ cross section data at approximately the same angles in a series of Pb isotopes 
by omitting isotopic differences in the transformation from the laboratory frame to the center-of-mass frame, 
which is negligibly small in this mass-number region.

\subsection{Isotopic difference of neutron radii and densities}

\begin{figure}[!h]
\includegraphics[width=7cm]{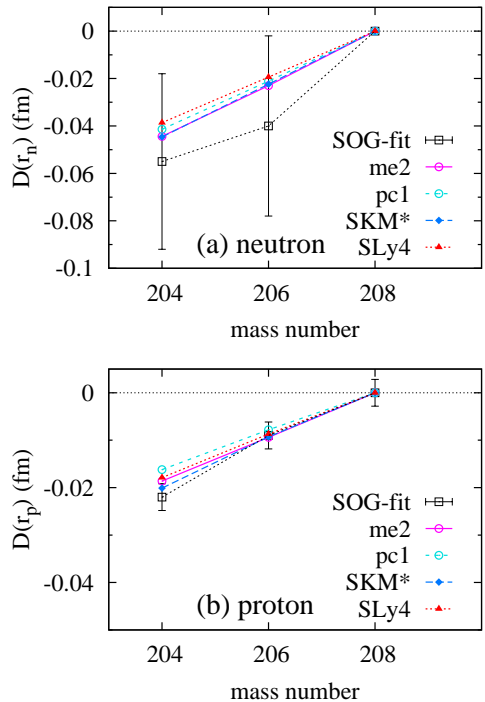}
  \caption{
Isotopic differences for (a) neutron and (b) proton radii, $D(r_n)$ and $D(r_p)$, for the 
RHB (me2 and pc1) and SHFB (SKM* and SLy4)  calculations, and 
the experimental SOG-fit density.
\label{fig:rmsr-d208}}
\end{figure}

The isotopic radius differences $D(r_n)$ and $D(r_p)$ are 
nothing but the relative radii of $\PbA$, measured from $\Pb208$.
The theoretical values of $D(r_n)$ and $D(r_p)$ are shown in Fig.~\ref{fig:rmsr-d208}
together with the experimental $D(r_n)$ of the SOG-fit density and the $D(r_p)$ values~\cite{Zenihiro:2010zz}.
The theoretical results of RHB (me2 and pc1) and SHFB (SKM* and SLy4) 
calculations are consistent  and show 
linear dependences of $D(r_n)$ and $D(r_p)$ upon the neutron-number difference $N-126$  
because of the isotopic systematics of $r_n$ and $r_p$ as discussed previously.
The theoretical values of the isotopic proton radius difference, $D(r_p)=-0.02$ fm for $\Pb204$ and $D(r_p)=-0.01$ fm for $\Pb206$, 
agree well with the experimental values reduced from the charge radii.
For the isotopic neutron radius difference, the theoretical values are  independent from structure models as 
$D(\Pb204;r_n)\approx -0.04$ fm and $D(\Pb206;r_n)\approx -0.02$ fm. 
They are contained in the experimental errors 
extracted from the $(p,p)$ data at 295 MeV, but deviate from the 
experimental best-fit values $D(r_n)= -0.055$~fm for $\Pb204$ and $D(r_n)= -0.040$~fm for $\Pb206$ 
of the SOG-fit neutron density. 

\begin{figure}[!h]
\includegraphics[width=6cm]{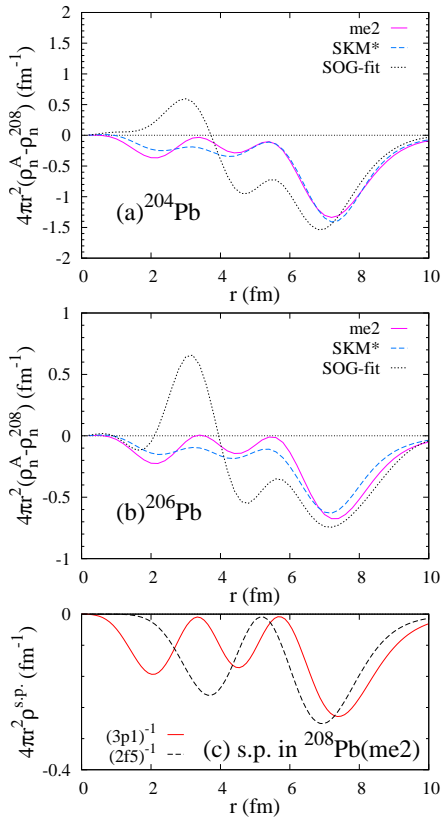}
  \caption{
Isotopic neutron-density differences ($D(\rho_n)$) of (a) $\Pb204$ and (b) $\Pb206$  
for the SOG-fit, me2, and SKM* densities. 
(c) One-neutron hole densities ($\rho^\textrm{s.p.}(r)$) of the $3p_{1/2}$ and $2f_{5/2}$ orbits in the $\Pb208$ core obtained by the RHB (me2) calculation. $4\pi r^2D(\rho_n)$ and $4\pi r^2\rho^\textrm{s.p.}(r)$ 
are plotted as functions of $r$.
\label{fig:dens-compare}}
\end{figure}

The isotopic neutron-density differences in $\Pb204$ and $\Pb206$  
of the SOG-fit, me2, and SKM* densities are shown in Fig.~\ref{fig:dens-compare}. 
The one-neutron hole densities of the $3p_{1/2}$ and $2f_{5/2}$ orbits in $\Pb208$ of the me2 density 
are shown for comparison. 
In the me2 and SKM* densities, $4\pi r^2D(\rho_n)$ shows 
a remarkable reduction of the surface density in the $6 \lesssim r  \lesssim 8$~fm region 
from $\Pb208$ to $\Pb206$ because of 
the dominant $(3p_{1/2})^{-2}$ and $(3p_{3/2})^{-2}$ contributions, but
contains the paring effect and other higher-order effects beyond perturbation such as $\Pb208$-core shrinkage. 
Compared with the enhanced amplitude of  $4\pi r^2D(\rho_n)$ at the surface, 
the amplitude in the internal region is relatively small and shows almost no specific character except for 
a small oscillation of the $3p$-hole component.
The SOG-fit density shows a $4\pi r^2D(\rho_n)$ surface amplitude  similar to the theoretical densities in 
the $6 \lesssim r  \lesssim 8$~fm region, but a strange behavior occurs
in the $2 \lesssim r  \lesssim 4$~fm region---that is, a sharp peak with an opposite sign. 
Since the proton elastic scattering at 295 MeV is insensitive to this internal region, 
this sharp peak may be an artifact 
that accidentally occurred in the SOG fitting under the condition of total neutron-number conservation. 
However, this artifact in the internal region causes a sudden increase of $r_n$ from $\Pb206$ to $\Pb208$
as understood by the relationship between $D(\rho_n)$ and $D(r_n)$ given in Eq.~\eqref{eq:dr-drho}.
Another difference between the SOG-fit and theoretical densities is seen in the  $4 \lesssim r \lesssim 6$~fm region.

\subsection{Isotopic cross-section ratio}

\begin{figure}[!h]
\includegraphics[width=0.5\textwidth]{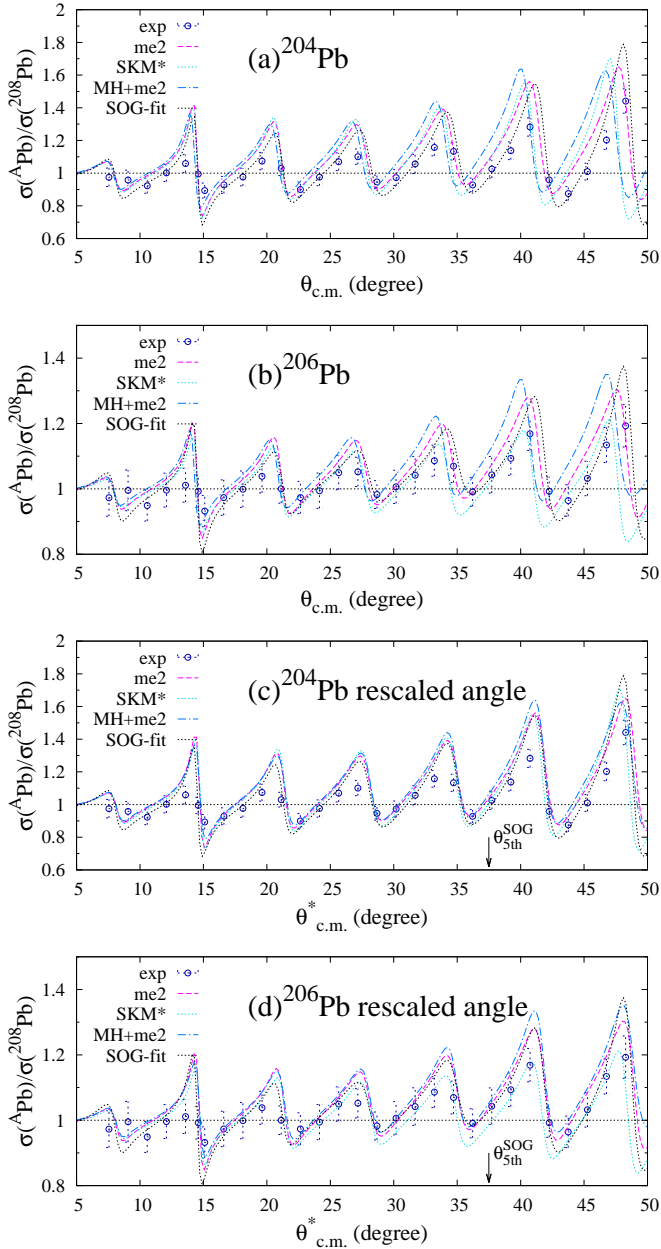}
  \caption{
Isotopic cross-section ratios $R(\sigma)$ of (a) $\Pb204(p,p)$ and (b) $\Pb206(p,p)$ to $\Pb208(p,p)$ at 295~MeV 
obtained with the 
ddMH model using the SOG-fit, me2, and SKM* densities plotted as functions of $\theta_\textrm{c.m.}$ together 
with the experimental values obtained from the $(p,p)$ cross section data~\cite{Zenihiro:2010zz}. 
The result obtained using the original MH model using the me2 (MH+me2) density
is also shown.
The $\theta^*_\textrm{c.m.}$-plots of $R(\sigma)$ for (c) $\Pb204$ and (d) $\Pb206$ for the rescaled angles 
$\theta^*_\textrm{c.m.}=(\theta^\textrm{SOG}_\textrm{5th}/\theta^\textrm{cal}_\textrm{5th})\theta_\textrm{c.m.}$ adjusted to fit the
fifth peak angles ($\theta^\textrm{theor}_\textrm{5th}$) of the $\Pb208(p,p)$ cross sections of theoretical results
to that ($\theta^\textrm{SOG}_\textrm{5th}$) of the SOG-fit result.
The fitting angle $\theta^\textrm{SOG}_\textrm{5th}$ is indicated using arrows. 
\label{fig:cross-compare-me2}}
\end{figure}

The isotopic cross-section ratios $R(\sigma)$ of $\Pb204$ and $\Pb206$ to $\Pb208$ are calculated by the ddMH model using 
the SOG-ft and theoretical densities. Figure \ref{fig:cross-compare-me2} compares the results obtained using 
the SOG-fit, me2, and SKM* densities, with the experimental values. 
The result calculated with the original MH model using the me2 density is also compared
to see the dependence on the effective $NN$ interaction in the reaction models. 
The isotopic cross-section ratio shows an oscillating behavior that reflects the diffraction pattern of the 
cross sections; the dip positions of $R(\sigma)$ correspond to the dip angles of the $\PbA(p,p)$ cross sections, whereas the peak positions of $R(\sigma)$ correspond to the dip angles of the $\Pb208(p,p)$ cross sections.
In principle, the oscillation amplitude of $R(\sigma)$ is sensitive to the size difference 
between isotopes. The size shrinkage of  $\PbA$ from $\Pb208$ expands 
the  diffraction pattern of $\sigma(\PbA)$, shifting the dip positions  slightly to backward angles
as compared with $\sigma(\Pb208)$. This shift that originates in the nuclear size shrinkage 
increases the amplitude in one cycle of the oscillation of $R(\sigma)$. 
Comparing different model calculations 
the oscillation interval of $R(\sigma)$ depends upon the structure and 
reaction models
because of the model dependence of the dip positions 
of $\sigma$ as shown in Fig.~\ref{fig:cross-ratio-pc1}; however, 
the oscillation amplitude of $R(\sigma)$ is similar between different models. 

To see more details of the oscillation amplitude, 
$R(\sigma)$ is plotted for yhr rescaled angles 
$\theta^*_\textrm{c.m.}=(\theta^\textrm{SOG}_\textrm{5th}/\theta^\textrm{cal}_\textrm{5th})\theta_\textrm{c.m.}$ 
so as to adjust the fifth peak angle ($\theta^\textrm{SOG}_\textrm{5th}$) of the $\Pb208(p,p)$ cross sections 
of SOG-fit with that ($\theta^\textrm{theor}_\textrm{5th}$) of the other calculations.
The $\theta^*_\textrm{c.m.}$-plots of $R(\sigma)$ for $\Pb204$ and $\Pb206$ are shown in Figs.~\ref{fig:cross-compare-me2}(c) and (d), respectively.
After rescaling the angles, 
the model dependence of  $R(\sigma)$ becomes small except for the $\theta^* > 42^\circ$ region 
as expected from that the isotopic neutron-density differences $D(\rho_n)$ in the surface region being similar 
between various theoretical densities and the SOG-fit density. 
This result indicates that
a model-independent discussion is possible in isotopic analysis
using the rescaled angles $\theta^*$.

Compared with the experimental $R(\sigma)$ obtained from the $(p,p)$ cross section  data, 
the $R(\sigma)$s calculated using the theoretical and SOG-fit densities 
slightly overshoot the experimental oscillation amplitude  for $\Pb204$ and $\Pb206$
as shown by the  $\theta^*$-plots in Figs.~\ref{fig:cross-compare-me2}(c) and (d), respectively.

\section{Hole-model analysis}\label{sec:analysis2}

\subsection{Model of the $\Pb208$ core with holes}
As discussed  previously, isotopic similarities are found in the neutron density and $(p,p)$ cross sections 
in a series of Pb isotopes, indicating that isotopic differences 
can be described by a perturbative treatment based on the $\Pb208$ core. 
By assuming a $\Pb208$ core and hole contributions,  
I introduce a model (called the hole model in this paper) for the neutron densities of $\Pb204$ and $\Pb206$ 
and discuss the connection between  $D(\rho_n)$ 
and $R(\sigma)$. 

The hole-model neutron densities of $\Pb204$ and $\Pb206$ are expressed using two parameters for 
the neutron-hole contribution and size scaling of the $\Pb208$-core as
\begin{align}
\rho_n(\PbA;r)&={\cal N}_0\left\{\rho_n(\Pb208;r/s)-\nhole\rho^\textrm{s.p.}_{3p_{1/2}}(r/s)\right\},\\
{\cal N}_0&=\frac{N}{126-\nhole}\frac{1}{s^3},\\
\delta_s &\equiv s-1
\end{align}
where ${\cal N}_0$ is the normalization factor for the total neutron number, 
$\rho^\textrm{s.p.}_{3p_{1/2}}$ is the neutron single-particle density of the $3p_{1/2}$ orbit in the 
$\Pb208$-core, $s$ is the $r$-scaling factor of the core-size shrinkage, which is
given by the scaling parameter $\delta_s$, 
and $\nhole$ is a model parameter for the $3p_{1/2}$-hole contribution. 
Note that $\nhole$ does not necessarily equal the actual neutron $3p_{1/2}$-hole number but is a parameter of the 
effective hole number for the $3p_{1/2}$-orbit contribution to neutron density. 
For the no-scaling~($\delta=0$) case, the hole model neutron density $\rho_n(\PbA;r)$ is approximately written as 
\begin{align}
&\rho_n(\PbA;r)\approx \rho_n(\Pb208;r)\nonumber\\
&-\nhole\rho^\textrm{s.p.}_{3p_{1/2}}(r)-(126-N-\nhole)\frac{\rho_n(\Pb208;r)}{126},
\end{align}
meaning that the total hole contribution is approximated by the $3p_{1/2}$-hole contribution 
and an overall reduction in the total neutron density, which contains contributions from other orbits.
With this expression of two parameters, $\nhole$ for the 
effective $3p_{1/2}$-hole number and $\delta_s$ for the size shrinkage, the hole model can simulate the densities of 
various configurations such as the $(2f_{5/2})^{-h}$ configuration and also describe 
``equivalent'' neutron densities to other theoretical densities that reproduce $R(\sigma)$. 
Examples of the hole model neutron densities equivalent to the $(2f_{5/2})^{-h}$ configuration and 
the me2 density are demonstrated in Appendix \ref{sec:app1}. 

In the present hole-model analysis, the hole-model density is used only for the neutron densities of $\Pb204$ and $\Pb206$ but 
the proton part is unchanged from the original proton densities of $\Pb204$ and $\Pb206$.
For the parameter $\nhole$ of the effective $3p_{1/2}$-hole number, 
the hole-model densities with $\nhole=0$, 0.5, $\ldots,$ 2 are used
and labeled as $(\textrm{0h},\delta_s)$, $(\textrm{0.5h},\delta_s)$, $\cdots$, 
$(\textrm{2h},\delta_s)$, respectively. 

I first perform  isotopic analysis using the 
hole model with the $\Pb208$-core and $3p_{1/2}$-orbit densities
obtained by the me2 calculation to clarify the correspondence of $R(\sigma)$ to  
$D(\rho_n)$ and $D(r_n)$. 
Then, I perform the hole-model analysis using the SOG-fit $\Pb208$-core and the me2 $3p_{1/2}$-orbit densities 
to obtain optimized parameter sets $h$ and $\delta_s$ of the hole model by fitting the experimental $R(\sigma)$ obtained 
by the Pb$(p,p)$ cross section data at 295~MeV. 

\subsection{Hole-model analysis with the me2-core density} \label{subsec:me2-core}
\begin{figure}[!h]
\includegraphics[width=0.5\textwidth]{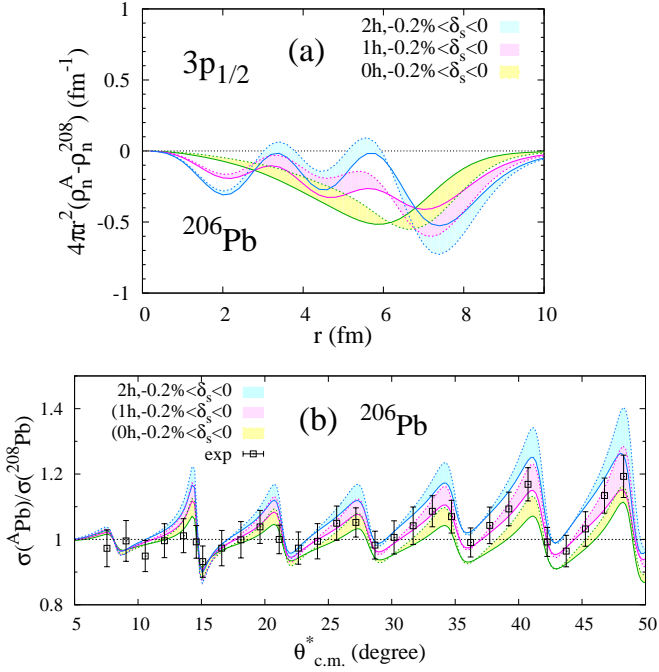}
  \caption{
(a) Isotopic neutron-density difference $D(\rho_n)$ and (b) isotopic cross-section ratio $R(\sigma)$ for $\Pb206$ 
obtained using the 
hole model (me2-core) with (0h,$-0.2\% \le\delta_s \le 0\%)$,  (1h,$ -0.2\% \le \delta_s \le 0\%)$, and  (2h,$ -0.2\% \le \delta_s \le 0\%)$, 
as shown by the color-filled areas surrounded by solid and dotted lines for $\delta_s=0\%$ and $\delta_s=-0.2\%$, respectively.
The calculated $R(\sigma)$ is plotted for rescaled angles $\theta^*_\textrm{c.m.}$.
The experimental values obtained from the $(p,p)$ data \cite{Zenihiro:2010zz} are also shown. 
\label{fig:dens-compare-me2-hole}}
\end{figure}

The isotopic neutron-density difference $D(\rho_n)$
and the isotopic cross-section ratio $R(\sigma)$ for $\Pb206$ obtained using the hole-model density with the me2-core 
are shown in Fig.~\ref{fig:dens-compare-me2-hole}. The 0h, 1h, and 2h results corresponding to the 
$3p_{1/2}$-hole numbers $\nhole=0$, 1, and 2 in the range of $-0.2\% < \delta_s < 0$ (0.2--0\% core shrinkage)
are illustrated by the yellow, pink, and blue-colored areas, respectively.
The oscillation 
amplitude of $R(\sigma)$ depends upon the size-scaling parameter $\delta_s$; it becomes larger 
as $\Pb206$ size shrinks from the $\Pb208$ core. The increase of $h$ (the effective $3p_{1/2}$-hole number) changes 
the surface behavior of $D(\rho_n)$; it reduces and enhances the surface amplitude of $4\pi r^2 D(\rho_n)$ 
in the $r\lesssim 7$~fm and
7$\lesssim r \lesssim$ regions, respectively, 
as expected from the hole density $\rho^\textrm{s.p.}_{3p_{1/2}}$ (Fig.~\ref{fig:dens-compare}(c)). 
This change in $4\pi r^2 D(\rho_n)$ increases $R(\sigma)$. 

\begin{figure}[!h]
\includegraphics[width=0.5\textwidth]{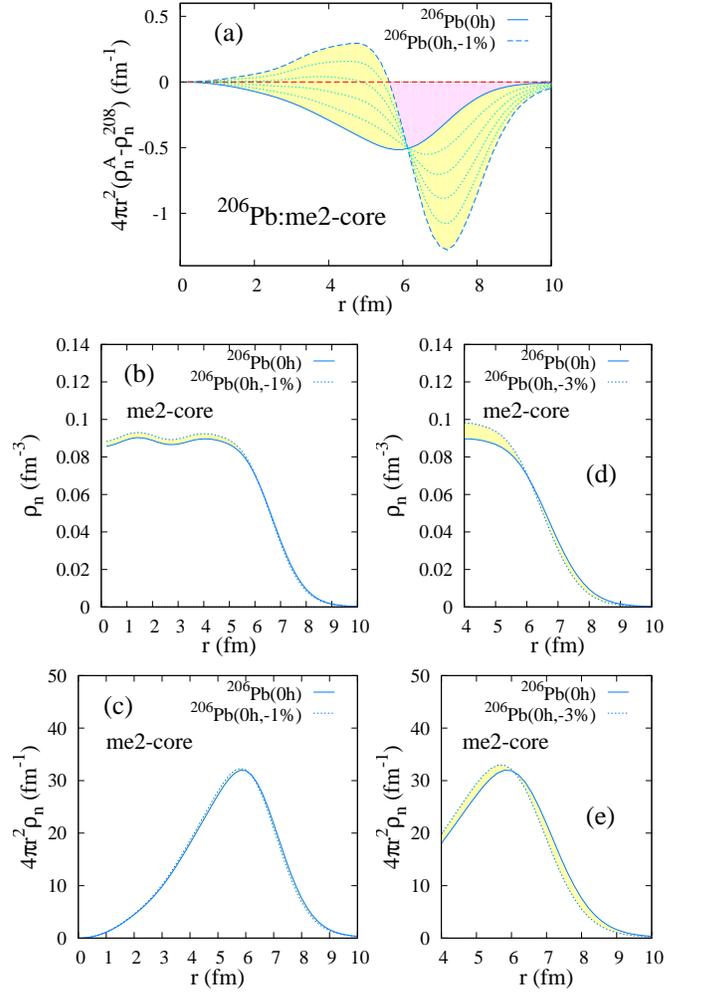}
  \caption{$\delta_s$ (size scaling) dependence of the hole-model neutron density of $\Pb206$ with the me2-core 
in the zero hole case, (0h,$\delta_s$).
(a) Isotopic neutron-density difference $D(\rho_n)$ for $\delta_s=\{-1\%,-0.8\%,\ldots,-0.2\%,0\%\}$. 
(b) $\rho_n(r)$ and (c) $4\pi r^2 \rho_n(r)$ for $\delta_s=-1\%$ and 
(d) $\rho_n(r)$ and (e) $4\pi r^2 \rho_n(r)$ for $\delta_s=-3\%$, as 
compared with the $\delta_s=0\%$~(no scaling) case. 
\label{fig:dens-me2-scale-demo}}
\end{figure}

\begin{figure}[!h]
\includegraphics[width=0.5\textwidth]{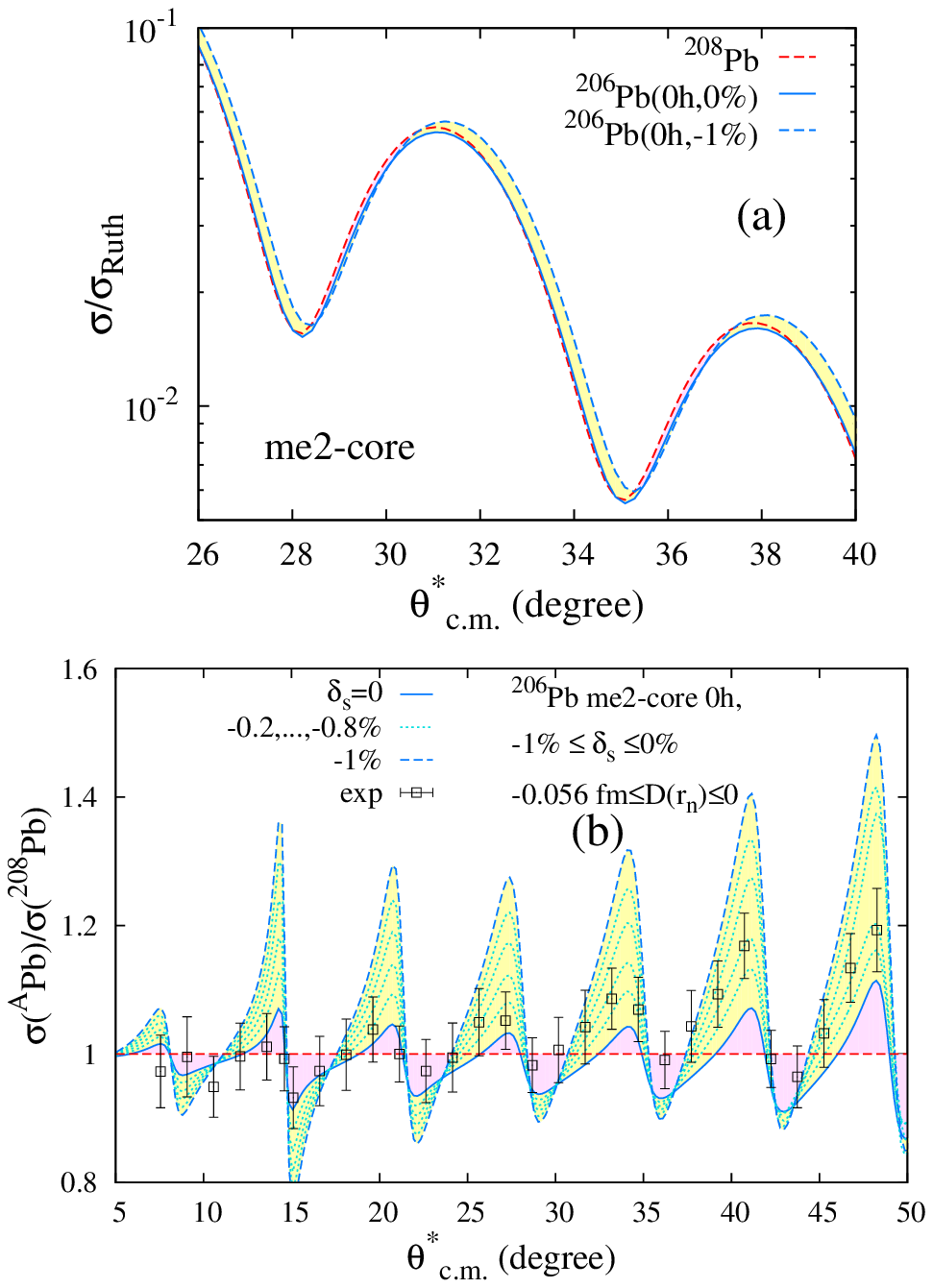}
  \caption{$\delta_s$~(size scaling) dependence of the cross sections and isotopic cross-section ratio 
for $\Pb206(p,p)$ at 295~MeV, as obtained 
using the hole model (me2-core) in the zero hole case, $(0h,\delta_s)$. 
(a) Rutherford ratio of the $\Pb206(p,p)$ cross sections for $\delta_s=0\%$ and $\delta_s=-1\%$ 
compared with that of $\Pb208(p,p)$.
(b) Isotopic cross-section ratio $R(\sigma)$ of $\Pb206$ to $\Pb208$,
as obtained using the (0h,$\delta_s$) densities with $\delta_s=\{-1\%,-0.8\%,\ldots,-0.2\%,0\%\}$, 
which give $D(r_n)=\{-0.56\textrm{~fm},-0.45\textrm{~fm}, \ldots,-0.11\textrm{~fm},0\textrm{~fm}\}$.
The calculated $R(\sigma)$ is plotted for rescaled angles $\theta^*_\textrm{c.m.}$.
The experimental values obtained from the $(p,p)$ data \cite{Zenihiro:2010zz} are also shown.
\label{fig:cross-model-me2}}
\end{figure}

Let me now discuss each effect of the size scaling and increase of $h$ upon
$R(\sigma)$ via changes in $D(\rho_n)$ and $D(r_n)$ in greater detail. 
First, I shall discuss the size-scaling effect 
by changing only $\delta_s$ of the 0h (no-hole) case of $\Pb206$. The results 
are shown in Figs.~\ref{fig:dens-me2-scale-demo} and \ref{fig:cross-model-me2}.
Figures~\ref{fig:dens-me2-scale-demo}(a), (b), and (c) show $4\pi r^2 D(\rho_n)$, 
$\rho_n(r)$, and $4\pi r^2 \rho_n(r)$ for $-1\%\le \delta_s\le 0\%$, respectively. Figures~\ref{fig:cross-model-me2}(a) and (b) 
show the results of $\sigma$ and $R(\sigma)$, respectively. 
For a clear visualization, $\rho_n(r)$ and $4\pi r^2 \rho_n(r)$ 
in an extreme case of $\delta_s=-3\%$ are shown in Figs.~\ref{fig:dens-me2-scale-demo}(d) and (e), respectively. 
As the size shrinks from $\delta_s=0\%$ to $-1\%$,  
$\rho_n(r)$  increases  in the $r < 6$~fm region and decreases in the $r > 6$~fm region 
(Figs.~\ref{fig:dens-me2-scale-demo}(b) and (d)); therefore, the 
peak position of $4\pi r^2 \rho_n(r)$ shifts inwardly (Figs.~\ref{fig:dens-me2-scale-demo}(c) and (e)). 
The inward shift of the $4\pi r^2 \rho_n(r)$ peak causes an outward shift 
of the diffraction pattern of the cross section (Fig.~\ref{fig:cross-model-me2}(a))
and enhances the oscillation amplitude of $R(\sigma)$, 
which monotonically increases as $\delta_s$ decreases
(Fig.~\ref{fig:cross-model-me2}(b)). This indicates that the oscillation amplitude of $R(\sigma)$ 
is a sensitive measure of the shrinkage of $\Pb206$ relative to $\Pb208$.

\begin{figure}[!h]
\includegraphics[width=0.5\textwidth]{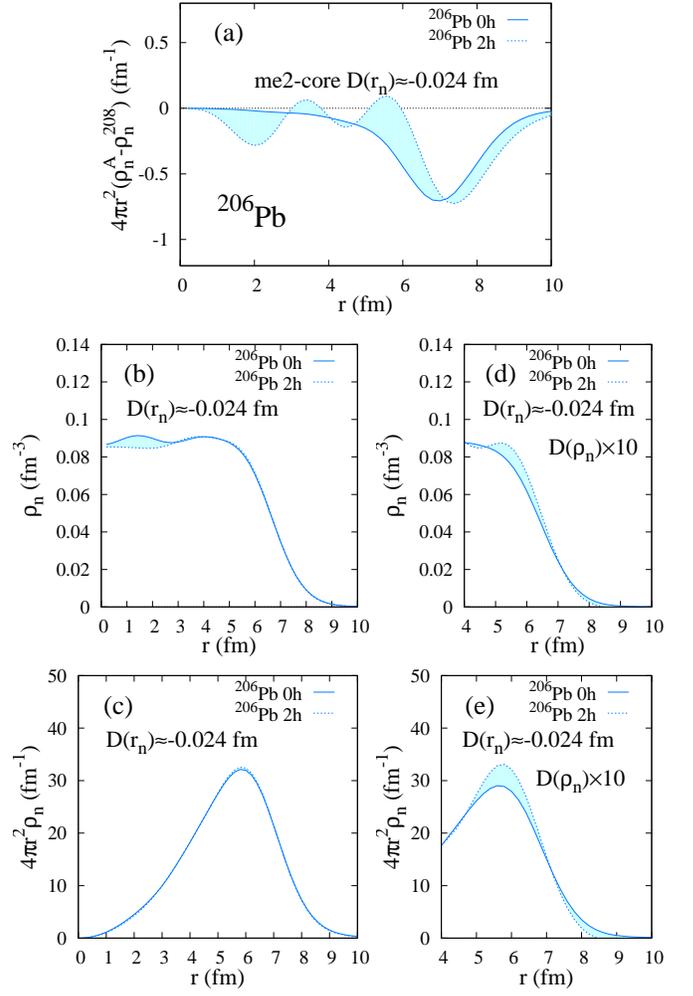}
  \caption{Neutron densities of $\Pb206$ for the hole model (me2-core) with 
(0h,$\delta_s=-0.4\%$) and (2h,$\delta_s=-0.2\%$) yielding $D(r_n)=-0.023$~fm and $-0.025$~fm, respectively. 
(a) Isotopic neutron-density difference $4\pi r^2D(\rho_n)$ and neutron densities 
(b) $\rho_n(r)$ and (c) $4\pi\rho_n(r)$. 
To see the difference between the 0h and 2h densities more clearly,  
$\rho_n(\Pb208)-10D(\rho_n)$ with 10 times enhanced $D(\rho_n)$ is displayed in (d),
and the $4\pi r^2$-weighted version is displayed in (e).
\label{fig:dens-me2-hole-demo}}
\end{figure}

\begin{figure}[!h]
\includegraphics[width=0.5\textwidth]{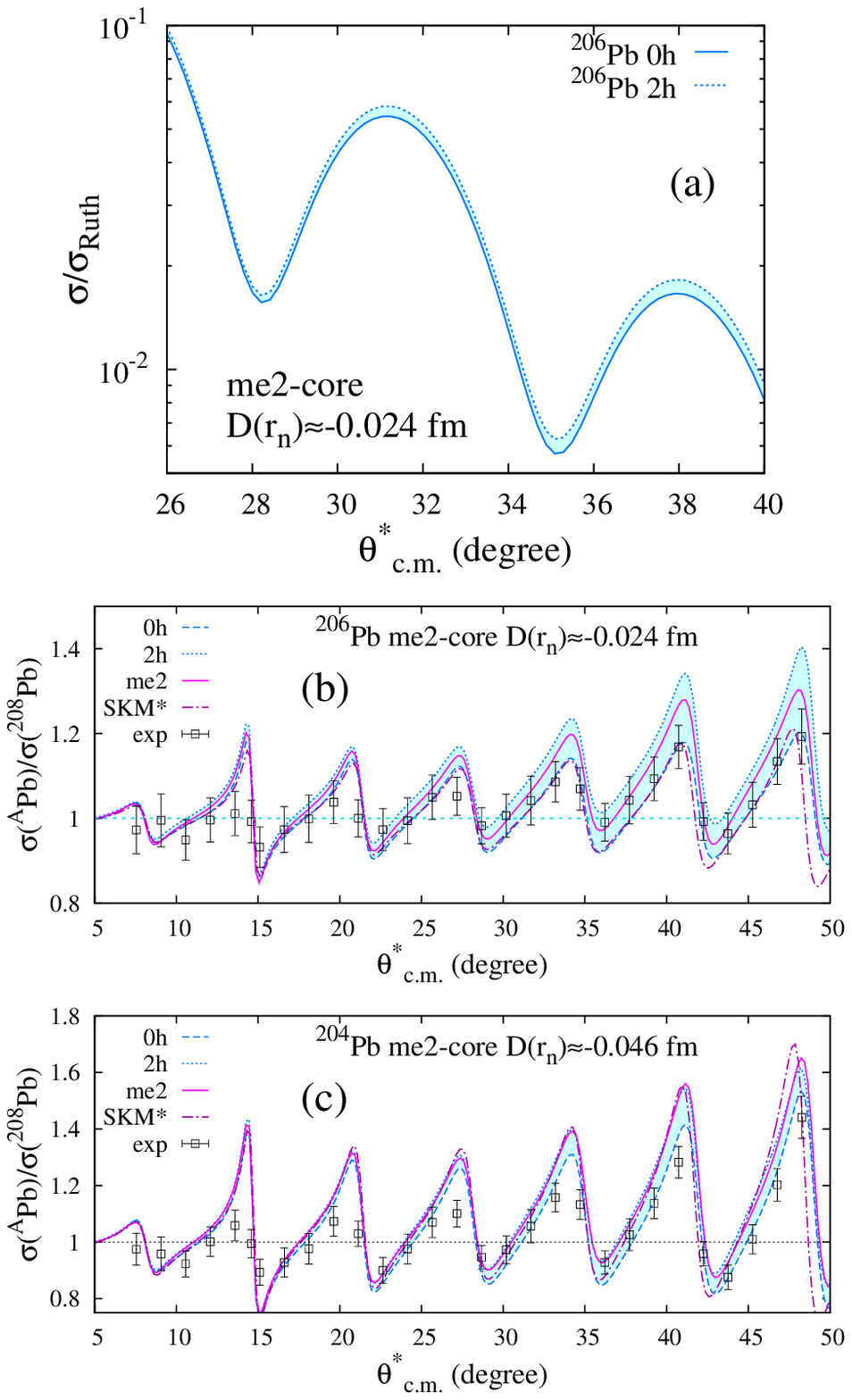}
  \caption{
(a) Rutherford ratio and (b) isotopic cross-section ratio $R(\sigma)$
for $\Pb206(p,p)$ at 295~MeV obtained using 
the hole model (me2-core) with (0h,$\delta_s=-0.4\%$) and (2h,$\delta_s=-0.2\%$), which 
respectively give $D(r_n)=-0.023$~fm and $-0.025$~fm.
(c) $R(\sigma)$ for $\Pb204$ obtained using the (0h,$\delta_s=-0.8\%$) density with $D(r_n)=-0.045$~fm and 
the (2h,$\delta_s=-0.6\%$) density with $D(r_n)=-0.047$~fm.
Theoretical $R(\sigma)$ values obtained using the me2 and SKM* densities and 
experimental values obtained from the 
$(p,p)$ data \cite{Zenihiro:2010zz} are also shown in (b) and (c). 
The calculated results are plotted for rescaled angles $\theta^*_\textrm{c.m.}$.
\label{fig:cross-model-me2-2}}
\end{figure}

Next, I discuss the $3p_{1/2}$-hole contribution effects, which provide a nontrivial change of
the density shape in the surface region. The density and cross sections obtained 
for the two-hole (2h) case are compared with those for the no-hole case in 
Figs.~\ref{fig:dens-me2-scale-demo} and \ref{fig:cross-model-me2-2}, respectively;
to eliminate the size changing effect, I choose $\delta_s$ independently 
for the 0h and 2h cases so that approximately the same $D(r_n)$ values will be given in the two cases as 
$D(r_n)\approx -0.024$~fm for $\Pb206$ and $D(r_n)\approx -0.046$~fm for $\Pb204$.  
The densities of $\Pb206$ are shown in Fig.~\ref{fig:dens-me2-scale-demo}, 
the results of the $\Pb206(p,p)$ cross sections are shown 
in Figs.~\ref{fig:dens-me2-scale-demo}(a) and (b), and 
those for $\Pb204(p,p)$ cross sections are presented 
in Fig.~\ref{fig:cross-model-me2-2}(c).
Figures~\ref{fig:dens-me2-scale-demo}(d) and (e) display 
$\rho_n(\Pb208)-10\times D(\rho_n)$ with 10 times enhanced $D(\rho_n)$ to show the 
difference between the 0h and 2h densities more clearly.
The dominant effect of the $3p_{1/2}$-hole contribution is an enhancement of the surface density
in the 5~fm$<r<$7~fm region  
around the peak position of $4\pi r^2\rho_n(r)$, 
whereas the contribution to the 
tail density in the $r >7$ ~fm region is relatively small (Fig.~\ref{fig:dens-me2-scale-demo}~(a),(c), and (e)). This density change by 
the $3p_{1/2}$-hole contribution
globally raises the cross sections of $\Pb206$ 
(Fig.~\ref{fig:cross-model-me2-2}(a)); therefore, it causes an upward shift of 
the isotopic cross-section ratio $R(\sigma)$ without changing its angular dependence (Fig.~\ref{fig:cross-model-me2-2}(b)). 
Note that, the parallel shift of $R(\sigma)$ with increasing $h$
is obtained only when $D(r_n)$ is kept constant.
The same analysis is performed for $\Pb204$, 
and a qualitatively consistent result is obtained (Fig.~\ref{fig:cross-model-me2-2}(c)).

In Figs.~\ref{fig:cross-model-me2-2}~(b) and (c), 
the $R(\sigma)$ obtained using the hole-model density is compared with that obtained using 
the me2 and SKM* densities. 
The me2(SKM*) density gives $D(r_n)=-0.023$~fm ($-0.022$~fm) for  $\Pb206$  and 
$-0.044$~fm ($-0.045$~fm) for $\Pb204$, which are consistent with the $D(r_n)$ values  
of the hole-model density presented herein. 
The oscillation amplitude of $R(\sigma)$ is almost consistent between the hole-model, me2, and SKM* densities, meaning that it 
is determined by $D(r_n)$ 
independently from details of model densities. 

The present analyses of $\delta_s$ and $h$ dependences indicate the clear correspondence of 
$R(\sigma)$ to $D(\rho_n)$ and $D(r_n)$. Namely, the oscillation amplitude of $R(\sigma)$ 
is a good measure to determine the isotopic neutron radius difference $D(r_n)$, 
whereas deviation of the center value of the oscillation
from $R=1$ is a sensitive probe to the surface shape of the isotopic neutron-density difference $D(\rho_n)$.

\subsection{Hole-model analysis with the experimental SOG-fit core density}

\begin{figure*}[!h]
\includegraphics[width=\textwidth]{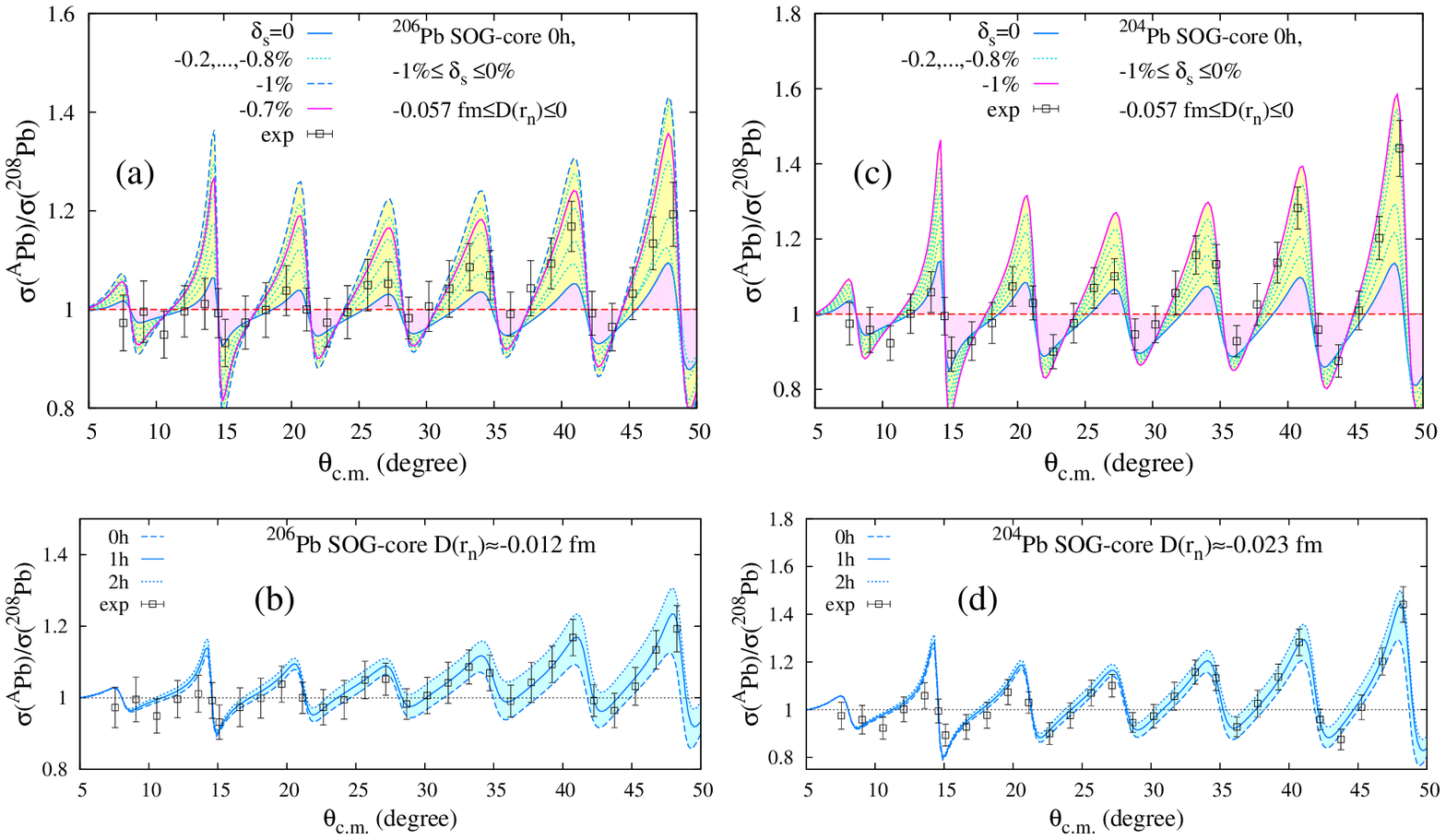}
  \caption{Isotopic cross-section ratio $R(\sigma)$ of $\PbA(p,p)$ to $\Pb208(p,p)$ at 295~MeV, 
as calculated using the hole model with the SOG-fit core.
Results for (0h,$\delta_s$) with $-1.0\%\le \delta_s \le 0\%$ for (a) $\Pb206$ and (c) $\Pb204$.
Red lines show the results obtained using the 
(0h,$-0.7\%$) hole model with $D(r_n)=-0.040$ fm of $\Pb206$ and (0h,$-1.0\%$) hole model with $D(r_n)=-0.057$ fm of $\Pb204$. 
(b) $R(\sigma)$ of $\Pb206$ obtained using 
the hole model with (0h,$\delta_s=-0.2\%$), (1h,$\delta_s=-0.1\%$), and (2h,$\delta_s=0\%$), which give 
$D(r_n)=-0.011$~fm, $-0.012$~fm, and $-0.012$~rm, respectively.
(d) $R(\sigma)$ of $\Pb204$ obtained using 
the hole model with (0h,$\delta_s=-0.4\%$), (1h,$\delta_s=-0.3\%$), and (2h,$\delta_s=-0.2\%$), which yield
$D(r_n)=-0.023$~fm.
The experimental values obtained from the $(p,p)$ data \cite{Zenihiro:2010zz} are also shown.
\label{fig:cross-compare-exp-hole}}
\end{figure*}

Using the SOG-fit core density, 
the hole-model analysis is performed in a similar way to Sec.~\ref{subsec:me2-core}
for the me2-core density, 
and qualitatively consistent results are obtained. 
The $R(\sigma)$ result obtained by the RIA+ddMH calculation with the hole-model~(SOG-core) density
is shown in Fig.~\ref{fig:cross-compare-exp-hole}. 
Figures ~\ref{fig:cross-compare-exp-hole}(a) and (b) show 
the $\delta_s$ and $h$ dependences of $R(\sigma)$ for $\Pb206$,
and Figs.~\ref{fig:cross-compare-exp-hole}(c) and (d) show the results for $\Pb204$.
For $\Pb206$, the hole-model density with (0h,$-0.7\%$) 
(which has $D(r_n)=-0.040$ fm, same as the SOG-fit density) overshoots 
the oscillation amplitude of the experimental $R(\sigma)$ (see the red line in  Fig.~\ref{fig:cross-compare-exp-hole}(a)). 
This indicates that 
the value $D(r_n)=-0.040$ fm of the SOG-fit density for $\Pb206$ may be too large;
similarly, the hole-model density of $\Pb204$ with (0h,$-1\%$) has 
$D(r_n)=-0.057$~fm, which is almost the same as the SOG-fit density, but 
seems to overshoot the oscillation amplitude of the experimental $R(\sigma)$
(see the red line in  Fig.~\ref{fig:cross-compare-exp-hole}(c)), meaning that 
the value of $D(r_n)=-0.057$ fm for $\Pb204$ is unlikely. 

The optimal parameter set of $h$ and $\delta_s$ is sought for 
the hole model to fit the  experimental $R(\sigma)$. 
First, I choose the favored parameter sets among the 0h densities,  
(0h,$-0.2\%$) with $D(r_n)=-0.011$ fm for 
$\Pb206$ and (0h,$-0.4\%$) with $D(r_n)=-0.023$ fm for $\Pb204$, 
to reproduce the slope in one cycle of the oscillation amplitude of the experimental $R(\sigma)$;
then, I change the effective hole number $h$, maintaining the optimal $D(r_n)$ values. 
The calculated $R(\sigma)$ obtained using the 
0h, 1h, and 2h densities with $D(r_n)\approx -0.01$~fm ($-0.02$~fm) for $\Pb206$ ($\Pb204$) is shown in 
Fig.~\ref{fig:cross-compare-exp-hole}(b) (Fig.~\ref{fig:cross-compare-exp-hole}(d)). Finally,
the (1h,$-0.01\%$) density with $D(r_n)=-0.012$~fm is obtained as an optimized solution for $\Pb206$,
which describes the experimental $R(\sigma)$, and an optimized set (1h,$-0.03\%$) with $D(r_n)=-0.023$~fm is 
obtained for $\Pb204$. The $R(\sigma)$ values calculated using the optimized solutions are 
shown by solid lines in Figs.~\ref{fig:cross-compare-exp-hole}~(b) and (d).

\subsection{Uncertainty in $D(r_n)$}

\begin{figure}[!h]
\includegraphics[width=0.5\textwidth]{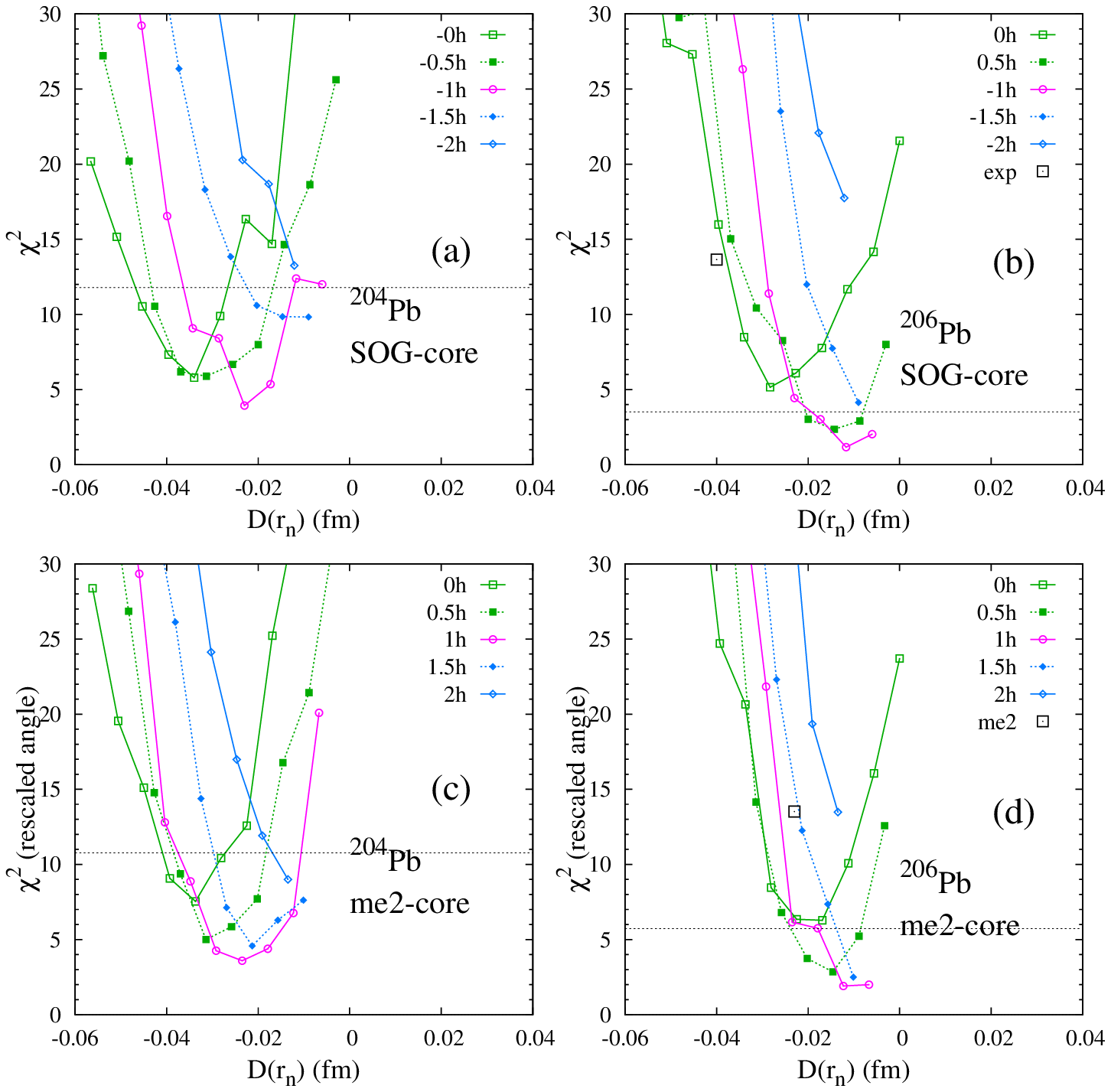}
  \caption{
$\chi^2$ of $R(\sigma)$ obtained by the ddMH calculation using the hole model with (0h,$\delta_s$), (0.5h,$\delta_s$), (1h,$\delta_s$), (1.5h,$\delta_s$), and (2h,$\delta_s$) for $\delta_s=\{0\%, 0.1\%, \cdots -1.0\%\}$
for (a) $\Pb204$ and (b) $\Pb206$ with the SOG-core and (c) $\Pb204$ and (d) $\Pb206$ with the me2-core.
The $\chi^2$ values are plotted as functions of $D(r_n)$. 
$\chi^2$ is calculated using 
17 datapoints of the $(p,p)$ data \cite{Zenihiro:2010zz}
excluding $\theta_\textrm{c.m.}\le 15.09^\circ$ at forward angles and 
$\theta_\textrm{c.m.}=21.13^\circ$, $27.17^\circ$, $28.67^\circ$, $34.71^\circ$, and  $42.24^\circ$ at dip angles. 
Rescaled angles $\theta^*_\textrm{c.m.}$ are used for the me2 and 
 hole model(me2-core) cases.
\label{fig:rmsr-chi}}
\end{figure}

To discuss the uncertainty in determining $D(r_n)$ from the experimental $R(\sigma)$,
I calculate the $\chi^2$ values of $R(\sigma)$ for the hole model. 
In the present analysis of $R(\sigma)$, the angular resolution of the experimental $(p,p)$ cross section data 
is not taken into account, but it 
can have crucial effects  at the forward and dip angles of the cross sections.
Therefore, $\chi^2$ is calculated using a total of 17 datapoints in a ``safe'' region by eliminating 
seven datapoints at the forward angles and five  at the dip angles.
In Fig.~\ref{fig:rmsr-chi}, 
the $\chi^2$ values obtained for $h=\{0,0.5,\ldots,2\}$ and  $\delta_s=\{0\%, 0.1\%, \cdots -1.0\%\}$ are 
plotted as functions of $D(r_n)$.

The results for $\Pb204$ and $\Pb206$ of the hole model with the SOG-fit core 
are shown in Figs.~\ref{fig:rmsr-chi}(a) and (b).
The absolute values of the calculated $\chi^2$ are meaningless
because the experimental errors of $R(\sigma)$ are estimated by assuming independent errors of the $\PbA$ and $\Pb208$
cross sections though the $A$-independent systematic errors should be removed.
To roughly estimate the uncertainty,  I here adopt
a criterion for the acceptable $\chi^2$ range as less than three times 
the minimum value $\chi^2_\textrm{min}$ as $\chi^2\lesssim 3 \chi^2_\textrm{min}$. 
From the acceptable ranges of this criterion, which are shown by dotted lines in the figures, 
$D(r_n)=-0.03$--$-0.006$~fm for $\Pb206$ and $D(r_n)=-0.05$--$-0.006$~fm for $\Pb204$ are obtained. 
The $\chi^2$ values for the SOG-fit density are $\chi^2=14$ for $\Pb206$ and $\chi^2=52$ for $\Pb204$, which 
exceed the acceptable range because the systematics between the Pb isotopes was not taken into 
account in the fitting of Ref.~\cite{Zenihiro:2010zz}.

For the hole model with the me2 core,  
the $\chi^2$ values are calculated using rescaled angles $\theta^*$, and 
the results obtained for $\Pb204$ and $\Pb206$ are shown in Figs.~\ref{fig:rmsr-chi}(c) and (d), respectively.
The results of the me2 core are almost consistent with those of the SOG-fit core
meaning that the present hole-model analysis of $R(\sigma)$ can obtain $D(r_n)$ values almost independently from the 
adopted $\Pb208$-core density.
In Table~\ref{tab:drn}, 
the values of $D(r_n)$ and $\chi^2$	obtained by the hole models (SOG-fit and me2 $\Pb208$ cores) are  
summarized in comparison with those of the SOG-fit and me2 densities of the Pb isotopes.


\begin{table}[!ht]
\caption{$\chi^2$ values of $R(\sigma)$ for the SOG-fit,
me2, hole model~(SOG-core), and hole model~(me2-core) densities. $D(r_n)$ values are also listed.
For the hole models, 
the $\chi^2$ and $D(r_n)$ values for the optimized parameter sets 
are listed together with the acceptable ranges estimated by $\chi^2$ values.
 \label{tab:drn}
}
\begin{center}
\begin{tabular}{lccccccccc}
\hline
\hline
	&	 \multicolumn{2}{c}{$\Pb204$}	&	 \multicolumn{2}{c}{$\Pb206$} \\		
	&	$D(r_n)$ (fm)	&	$\chi^2$	&	$D(r_n)$ (fm)	&	$\chi^2$	\\
SOG-fit	&	$-0.055$	&	52	&	$-0.040$	&	14	\\
me2	&	$-0.044$	&	52	&	$-0.023$	&	14	\\
	&		&		&		&		\\
hole model 	&	$D(r_n)$ (fm)	&	$\chi^2_\textrm{min}$	&	$D(r_n)$ (fm)	&	$\chi^2_\textrm{min}$	\\
 (optimized)	&\\
SOG-core	&	$-0.023$	&	1.2	&	$-0.012$	&	3.9	\\
me2-core	&	$-0.024$	&	1.9	&	$-0.012$	&	3.6	\\
(acceptable) &	$-0.05\sim -0.006$ 	&		&	$-0.03\sim -0.006$	&		\\
\hline
\hline
\end{tabular}
\end{center}
\end{table}

\section{Improved neutron densities and radii of Pb isotopes}\label{sec:ndens}

\subsection{Reconstruction of neutron densities}
\begin{figure}[!htp]
\includegraphics[width=0.5\textwidth]{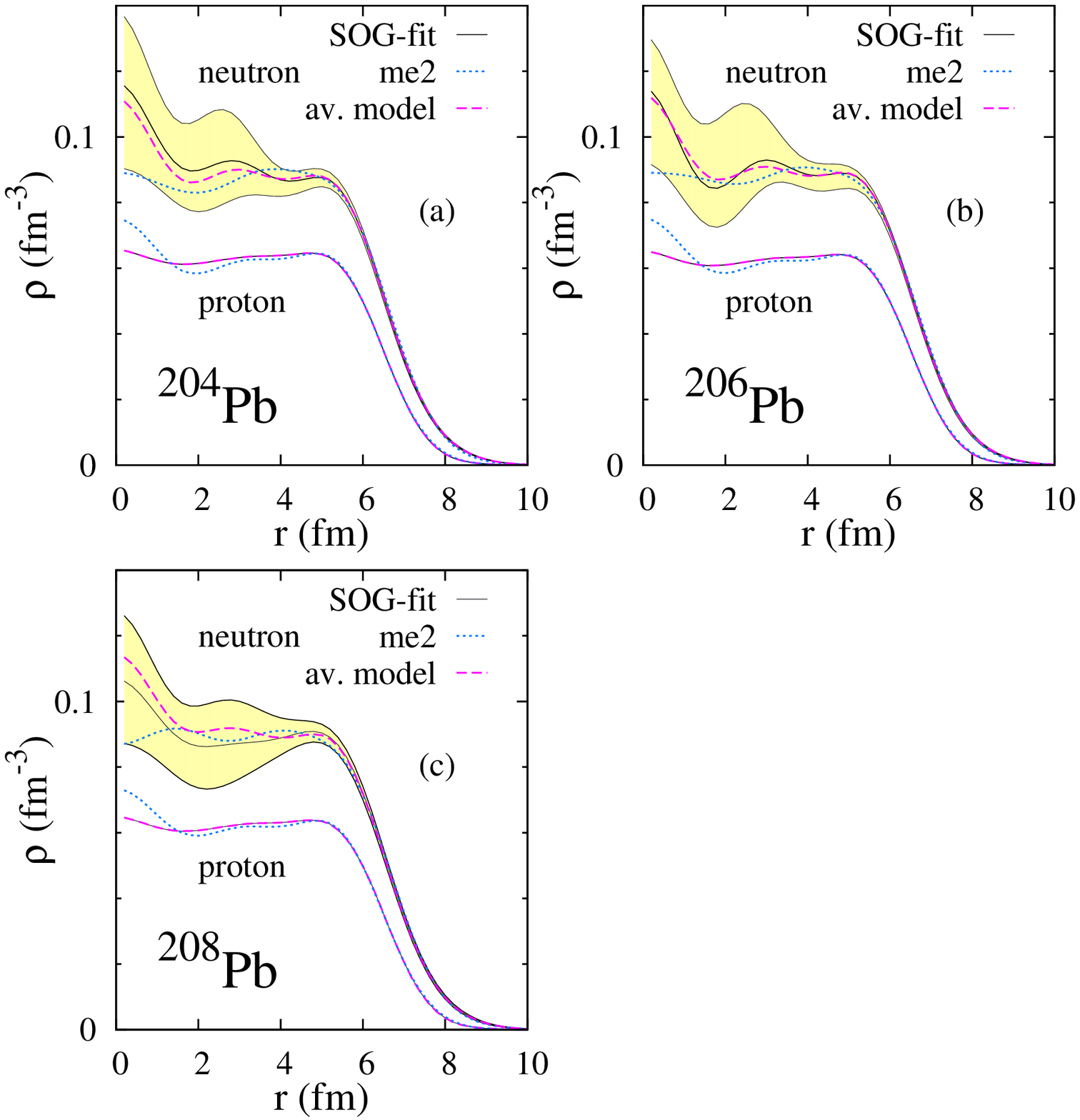}
  \caption{Same as Fig.~\ref{fig:dens-pn-1} but for the averaged-model densities compared with the 
me2 and SOG-fit densities. 
\label{fig:dens-pn-av}}
\end{figure}

\begin{figure}[!htp]
\includegraphics[width=6 cm]{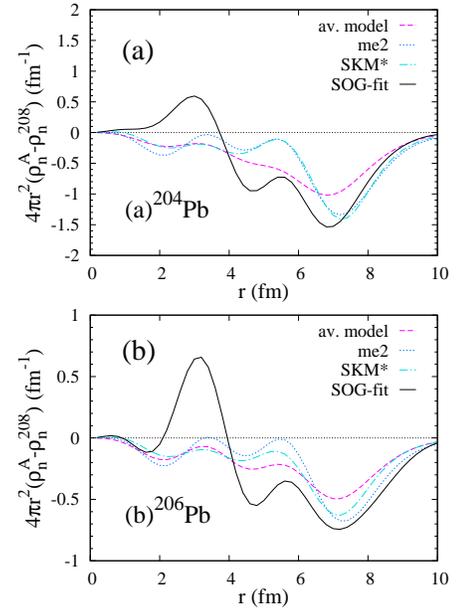}
  \caption{$4\pi r^2 D(\rho_n)$ for (a) $\Pb204$ and (b) $\Pb206$
of the isotopic neutron-density difference of the averaged-model densities compared with results of the 
SOG-fit, me2, and SKM* densities. 
\label{fig:dens-compare-av}}
\end{figure}

As mentioned previously, the SOG-fit density overshoots the oscillation 
amplitude of the experimental $R(\sigma)$. The reason for this failure in reproducing $R(\sigma)$ 
is that the fitting was performed independently for each isotope
without taking the isotopic systematics  into account. 
From the SOG-fit density,  
I reconstruct the improved neutron densities of Pb isotopes that can describe experimental $R(\sigma)$
as follows. 
First,  the neutron density of $\Pb206$ is obtained by 
averaging three densities of $\Pb206$, $\Pb206$, and $\Pb206$ given by the SOG-fit density 
to avoid a risk from uncertainty of the SOG-fit density in the internal region.
Next, the neutron densities of $\Pb208$ and $\Pb204$ are constructed to reproduce 
the $D(\rho_n)$ of the best solution of the hole model.
The density of the Pb isotopes obtained with these procedures is called the ``averaged-model'' density. 
In Fig.~\ref{fig:dens-pn-av}, the averaged-model density is shown 
in comparison with the SOG-fit and me2 densities. 
The averaged-model density is similar to the SOG-fit density.  
The $D(\rho_n)$s of the averaged model, SOG-fit, and me2 densities are 
compared in Fig.~\ref{fig:dens-compare-av}. 
Note that the $D(\rho_n)$ of the averaged model is tuned to fit the $R(\sigma)$ data in the hole-model analysis. 
The surface peak shape of $4\pi r^2 D(\rho_n)$ in the $6~\textrm{fm} \lesssim r\lesssim 8~\textrm{fm}$ region is similar between three kinds of densities; 
however, a difference is found in the internal $r\lesssim 5~\textrm{fm}$ region, to which $(p,p)$ at 295~MeV 
is insensitive.
The flopping behavior of $4\pi r^2 D(\rho_n)$ found in the SOG-fit density disappears 
in the averaged-model density, which shows a smooth behavior in the internal region similar to theoretical densities
such as the SKM* density, rather than the SOG-fit density.

\begin{figure}[!htp]
\includegraphics[width=0.5\textwidth]{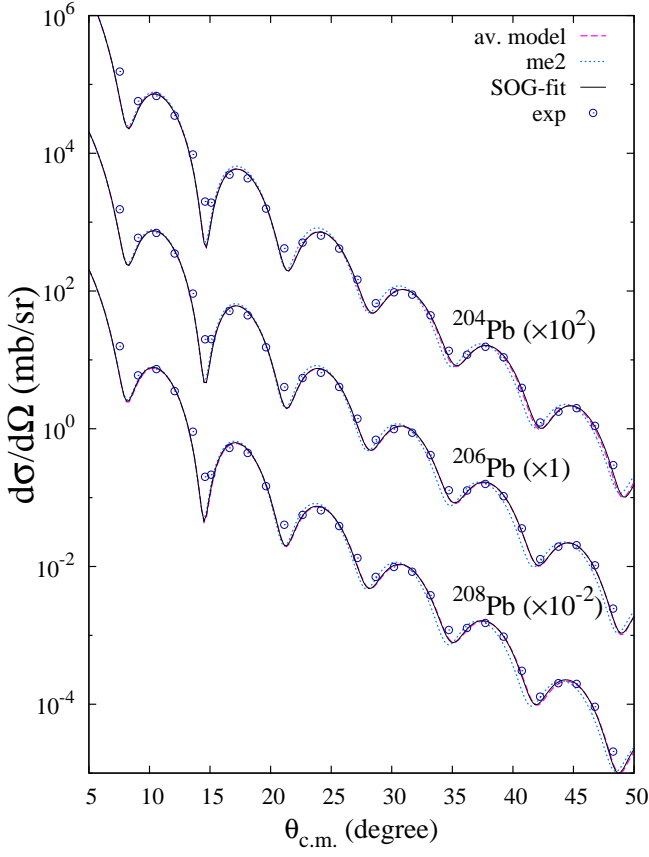}
  \caption{
Cross sections of $\textrm{Pb}(p,p)$ at 295 MeV obtained via RIA with the ddMH model 
using the present averaged model, SOG-fit, and the me2 densities of Pb isotopes,
 together with the experimental data~\cite{Zenihiro:2010zz}. 
\label{fig:cross-av}}
\end{figure}

\begin{figure}[!htp]
\includegraphics[width=0.5\textwidth]{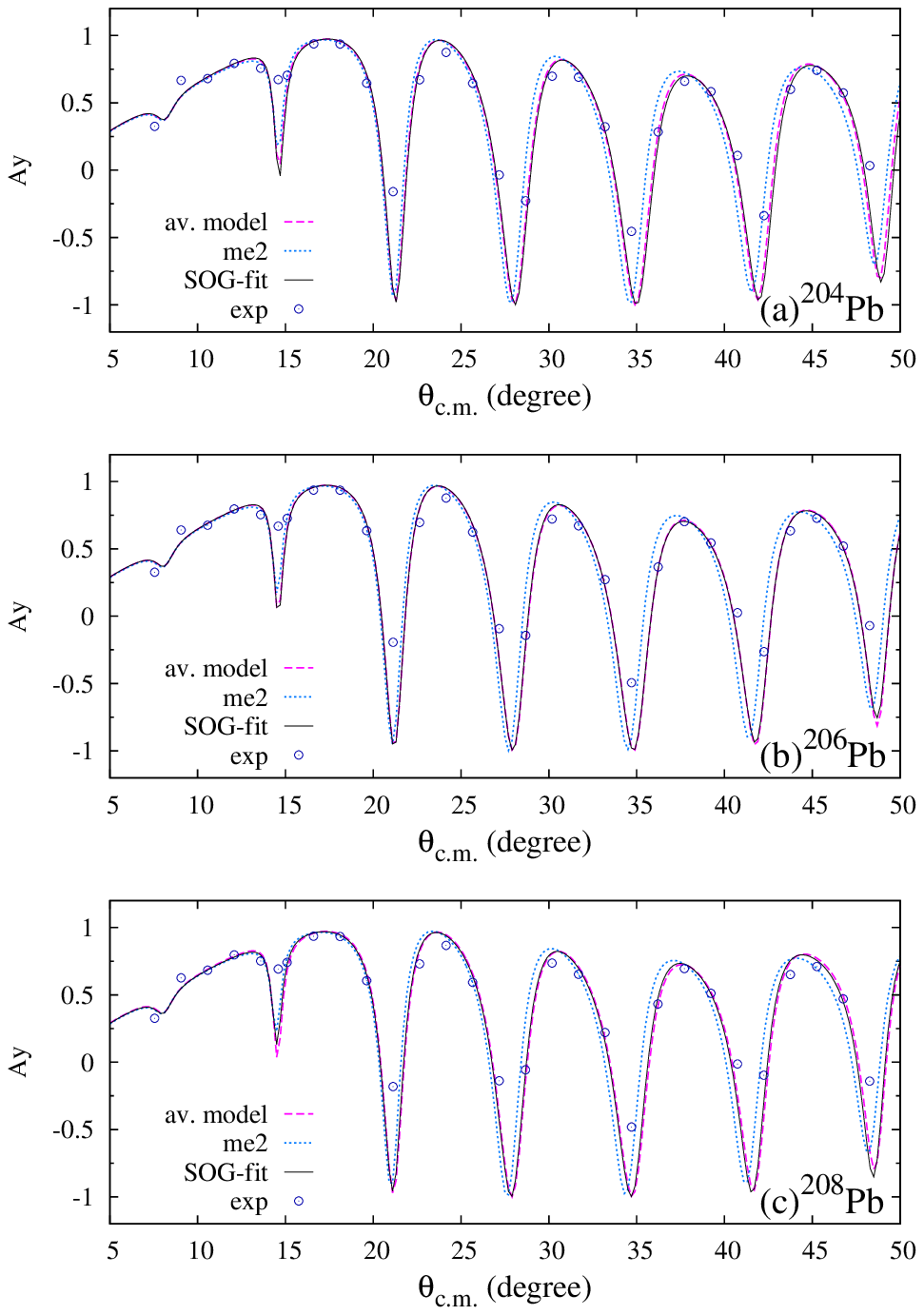}
  \caption{
Analyzing powers of Pb$(p,p)$ at 295 MeV obtained by RIA with the ddMH model using the present averaged model, SOG-fit, and me2 densities of Pb isotopes, together with the experimental data from~\cite{Zenihiro:2010zz}.
\label{fig:Ay-av}}
\end{figure}

\begin{figure}[!htp]
\includegraphics[width=0.5\textwidth]{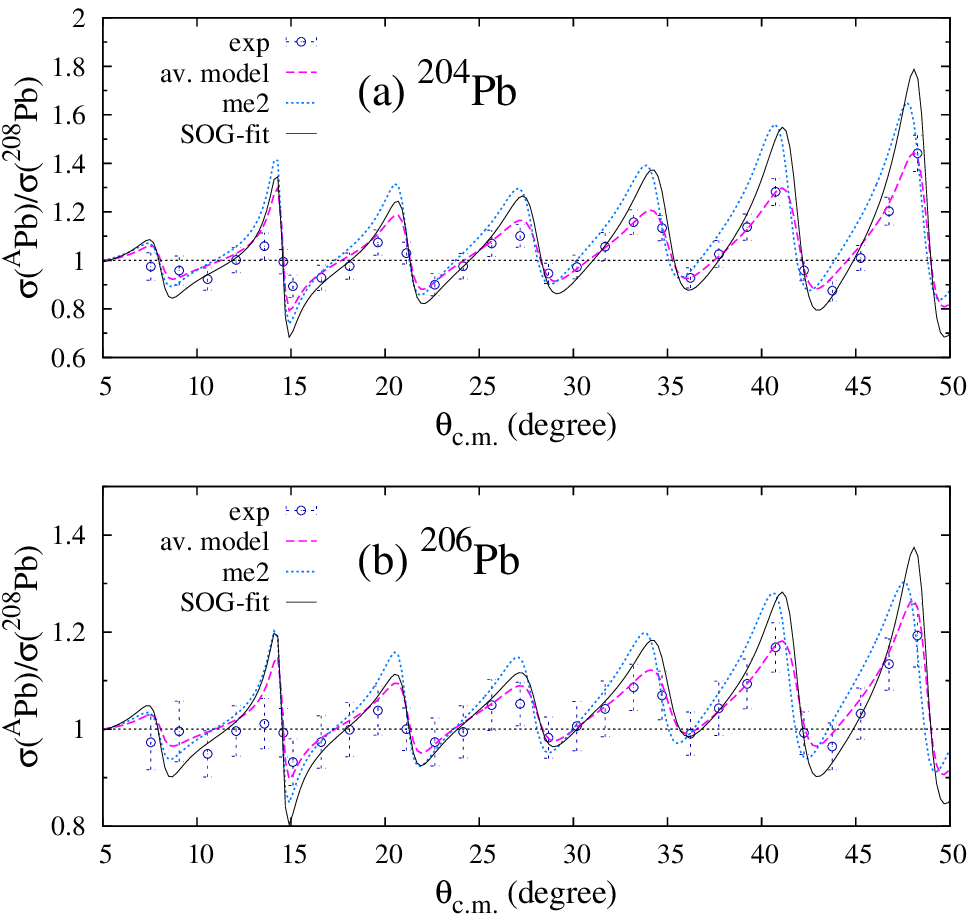}
  \caption{
Isotopic cross-section ratios $R(\sigma)$ of (a) $\Pb204(p,p)$ and (b) $\Pb206(p,p)$ to 
$\Pb208(p,p)$ at 295~MeV calculated using the present averaged model, SOG-fit, and me2 densities,
together with the experimental values obtained from the $(p,p)$ data~\cite{Zenihiro:2010zz}.
\label{fig:cross-compare-av}}
\end{figure}

Pb$(p,p)$ at 295~MeV is calculated using RIA+ddMH with the 
averaged-model density. The results of $R(\sigma)$, $\sigma$, and $A_y$ are shown 
in Figs.~\ref{fig:cross-compare-av}, \ref{fig:cross-av}, and \ref{fig:Ay-av}, respectively, 
and compared with those obtained with the SOG-fit and me2 densities. 
$R(\sigma)$ is significantly improved by the averaged-model density compared with other densities
because $D(\rho_n)$ is tuned to fit the $R(\sigma)$ data.
Moreover, the averaged-model density successfully 
describes the cross sections and analyzing powers in a quality almost equivalent to the SOG-fit density. 

\begin{figure}[!htp]
\includegraphics[width=7cm]{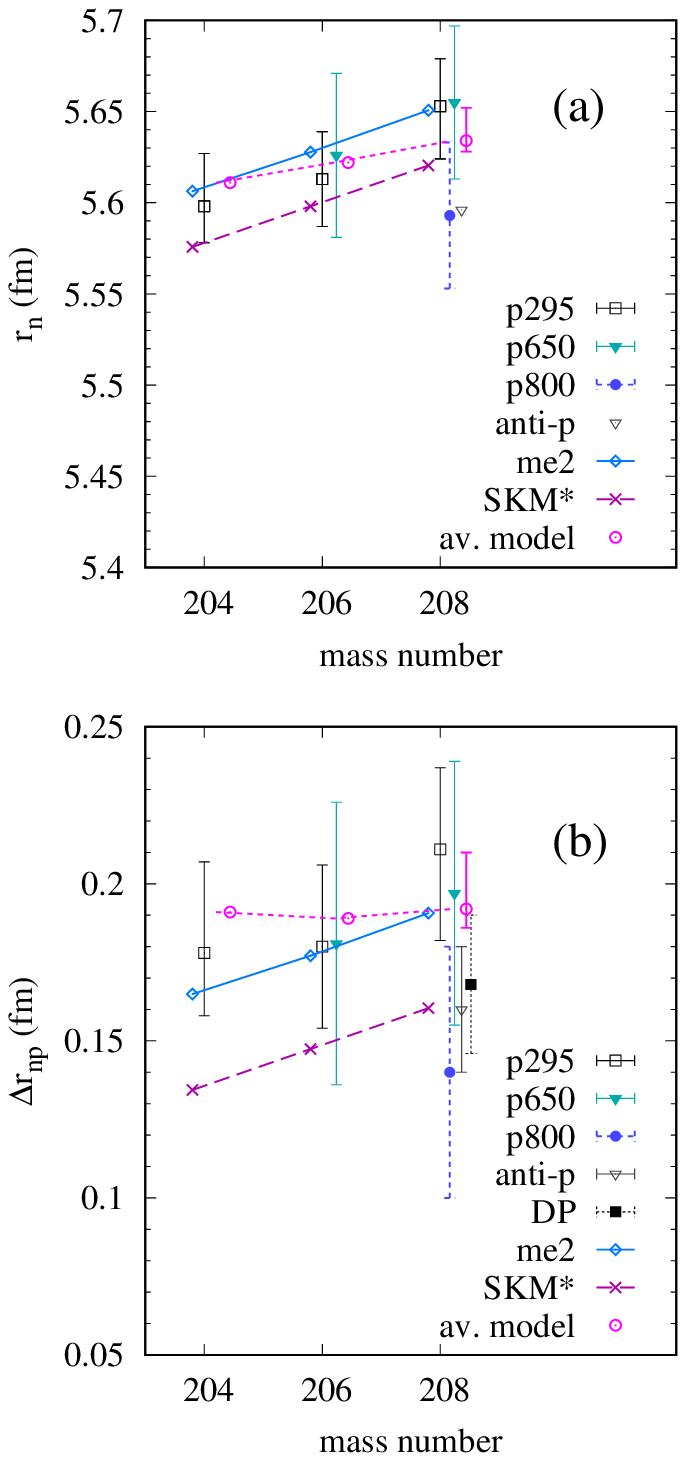}
  \caption{Same as Fig.~\ref{fig:rmsr} but for the present result obtained by the averaged model (open circles) 
in comparison with the theoretical and experimental values. 
For $\Pb208$, the error range of the averaged model is roughly estimated 
using the acceptable range $D(r_n)=-0.03\sim -0.006$~fm obtained via the hole-model analysis
and the $r_n$ value of $\Pb206$ of the averaged model.
\label{fig:rmsr-av}}
\end{figure}

In Fig.~\ref{fig:rmsr-av},
the neutron radius ($r_n$) and skin thickness ($\Delta r_{np}$) obtained by the present averaged model 
are shown in comparison with the experimental and theoretical values.
The averaged model yields smooth changes of $r_n$ and $\Delta r_{np}$ in a series of Pb isotopes from $\Pb204$ to 
$\Pb208$. The values are within the experimental errors of the SOG fitting extracted from the $(p,p)$ data at 295 MeV.
It is difficult to quantitatively discuss the systematic errors of the present result
because the angular resolutions and systematic errors of the experimental data are not 
taken into account in the present analysis. 
In Fig.~\ref{fig:rmsr-av},
I present a rough estimation of the error range of $r_n$ for $\Pb208$ 
using the acceptable range $D(r_n)=-0.03\sim -0.006$~fm obtained by hole-model analysis
and the $r_n$ value of $\Pb206$ of the averaged model.


\section{Summary}\label{sec:summary}

A new method of reaction analysis for proton elastic scattering was proposed 
by combining systematic analyses of nuclear structure and reaction in a series of isotopes. 
This method was applied to $\Pb204(p,p)$, $\Pb206(p,p)$, and $\Pb208(p,p)$ at $E_p=295$~MeV to obtain
improved neutron densities and radii from the experimental cross section data.
The reaction calculation of Pb$(p,p)$ at $E_p=295$~MeV was performed using RIA with the
effective $NN$ interaction of the ddMH model. 
For the target Pb density, the theoretical densities of 
the RHB (me2 and pc1) and SHFB (SKM* and SLy4)  calculations of spherical nuclei
and the experimental SOG-fit density are used as inputs of the reaction calculations.

The isotopic differences of the neutron density $D(\rho_n)$ and radius $D(r_n)$ of $\PbA$
from the reference $\Pb208$ were investigated, and 
the isotopic ratio $R(\sigma)$ of $\PbA(p,p)$ to the $\Pb208(p,p)$ cross sections was analyzed. 
The cross sections are sensitive to the profile of the surface-neutron density but
insensitive to the internal density. In the analysis of $D(\rho_n)$, 
the SOG-fit density was found to have a flapping behavior of $4\pi r^2 D(\rho_n)$ in the internal region,
causing an artificial increase of the neutron radius of $\Pb208$ that was inconsistent with theoretical predictions.

A further detailed analysis was performed with the hole model by assuming a  $\Pb208$ core and neutron-hole contributions
for the neutron densities of $\Pb204$ and $\Pb206$.
A clear correspondence between the surface-neutron density and the $(p,p)$ cross sections was clarified;
the oscillation amplitudes of $R(\sigma)$ is determined by the isotopic neutron radius difference $D(r_n)$, 
whereas the central values of the oscillation of $R(\sigma)$ are sensitive to the surface-neutron 
density around the peak position of $4\pi r^2 \rho_n(r)$.
Because of this correspondence between the isotopic structure differences ($D(\rho_n)$ and $D(r_n)$) and 
the isotopic cross-section ratio $R(\sigma)$, 
the $D(\rho_n)$ and $D(r_n)$ values 
can be safely extracted from the observed $(p,p)$ data with less model dependence.
By fitting the experimental $R(\sigma)$ with the hole model,
$D(r_n)=-0.012$~fm with the acceptable range $D(r_n)=-0.03 \sim -0.006$~fm was obtained for $\Pb206$.
Furthermore, the improved neutron densities of $\Pb204$, $\Pb206$, and $\Pb208$ were reconstructed from 
the SOG-fit density using the $D(\rho_n)$ obtained by hole-model analysis. 
The improved neutron densities of Pb isotopes successfully reproduced the experimental Pb$(p,p)$ data measured at 295~MeV,
including the isotopic cross-section ratio and the cross sections and analyzing powers. 
The results for $r_n$ and $\Delta r_{np}$ obtained by the improved neutron densities are 
reasonable values within the experimental errors and 
show smooth changes in a series of Pb isotopes from $\Pb204$ to $\Pb208$.

It should be noted that the present results for $D(r_n)$ and $D(\rho_n)$ can be improved by a further 
precise analysis by taking into account experimental errors in the angular resolution and 
the possible reduction of systematic errors in $R(\sigma)$, which were omitted in the present analysis.
Reanalysis of the Pb$(p,p)$ data at 295~MeV including isotopic systematics such as the $R(\sigma)$ data
is requested to extract a revised SOG-fit density from the $(p,p)$ data at 295~MeV.

The model uncertainties in the structure and reaction calculations were discussed 
by comparing the results obtained by RIA calculation with the ddMH and MH effective $NN$ interactions using various theoretical densities.
The present method of isotopic analysis was shown to be a useful tool for extracting neutron densities and radii
from the $(p,p)$ cross sections in a series of isotopes with less systematic uncertainties from model dependences.
The method can be extended straightforwardly to systematic analyses of neighboring nuclei, particularly 
for the $(p,p)$ data measured experimentally with the same setup at the same facility, because 
systematic errors can be significantly reduced in the experimental data of the isotopic cross-section ratio. 
This can be a great advantage for determining the nuclear sizes of unstable nuclei from $(p,p)$ data measured in inverse kinematics.

\begin{acknowledgments}
This work was motivated by the recent work of Ms. Shiyo Enyo. in her master thesis.
It was supported
by Grants-in-Aid of the Japan Society for the Promotion of Science (Grant Nos. JP18K03617 and 18H05407)
and by a grant of the joint research project of the Research Center for Nuclear Physics at Osaka
University.
\end{acknowledgments}

\clearpage
\appendix
\section{Equivalent $D(\rho_n)$ for the $(2f_{5/2})$-hole model and the me2 densities}\label{sec:app1}
\begin{figure*}[!h]
\includegraphics[width=15cm]{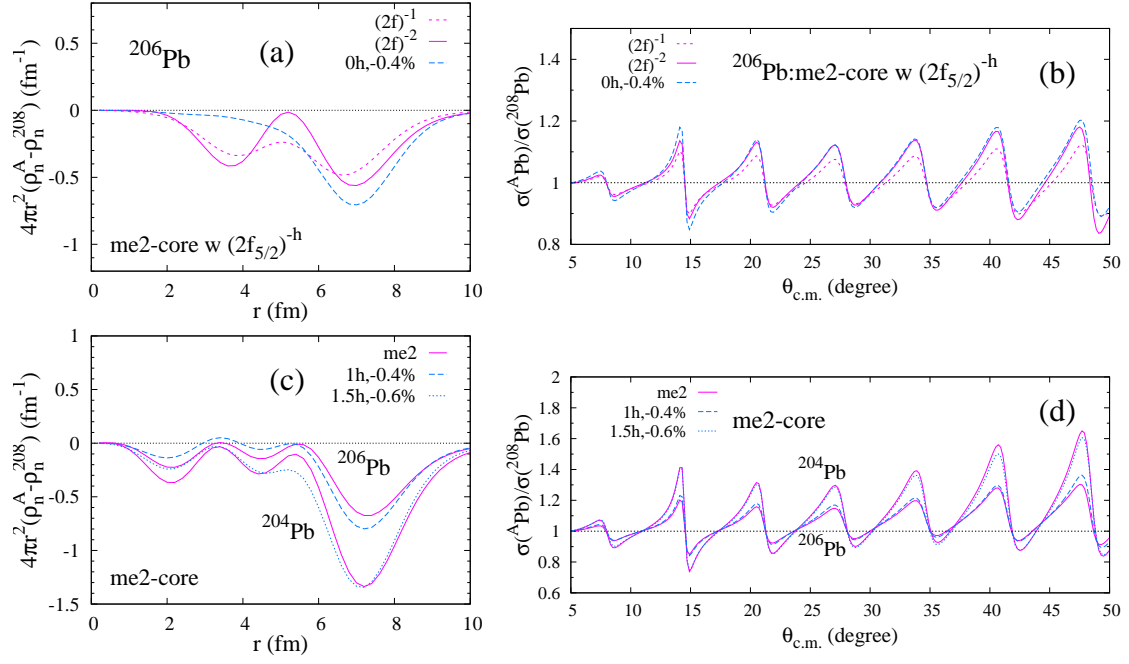}
  \caption{
(a) $D(\rho_n)$ and (b) $R(\sigma)$ obtained by the RIA+ddMH calculation using the hole model of the me2-core 
with $(2f_{5/2})^{-1}$ and $(2f_{5/2})^{-2}$ configurations (no scaling, $\delta_s=0$).
The results obtained using the equivalent $(3p_{1/2})$-hole-model density 
with (0h,$\delta_s=-0.4\%)$ that reproduces 
$R(\sigma)$ of the $(2f_{5/2})^{-2}$ configuration
are shown for comparison.
(c) $D(\rho_n)$ and (d) $R(\sigma)$ for the equivalent $(3p_{1/2})$-hole-model density,
together with those for the me2-density. 
The adjusted parameters for the  equivalent $(3p_{1/2})$-hole-model density
are (1h,$\delta_s=-0.4\%$) for $\Pb206$ and 
(1.5h,$\delta_s=-0.6\%$) for $\Pb204$. 
\label{fig:equivalent}}
\end{figure*}

$D(\rho_n)$ and $R(\sigma)$ obtained  using the hole model of the $(2f_{5/2})$-hole configurations
with the me2-core are shown in Fig.~\ref{fig:equivalent}~(a) and (b), respectively. 
The equivalent $(3p_{1/2})$-hole-model density,
which yields an approximately consistent result of $R(\sigma)$ with the $(2f_{5/2})$-hole model
is obtained by tuning the parameters $h$ and $\delta_s$ of the $(3p_{1/2})$-hole model.
The results obtained using the equivalent $(3p_{1/2})$-hole-model density adjusted 
to the hole model with the $(2f_{5/2})^{-2}$ configuration
are compared,
In Figs.~(c) and (d), the $D(\rho_n)$ and $R(\sigma)$  
obtained using the equivalent $(3p_{1/2})$-hole-model density adjusted to the me2-density
are compared with those of the me2-density.

\bibliographystyle{apsrev4-1}
\bibliography{RIA-pb-refs}

\begin{thebibliography}{33}%
\makeatletter
\providecommand \@ifxundefined [1]{%
 \@ifx{#1\undefined}
}%
\providecommand \@ifnum [1]{%
 \ifnum #1\expandafter \@firstoftwo
 \else \expandafter \@secondoftwo
 \fi
}%
\providecommand \@ifx [1]{%
 \ifx #1\expandafter \@firstoftwo
 \else \expandafter \@secondoftwo
 \fi
}%
\providecommand \natexlab [1]{#1}%
\providecommand \enquote  [1]{``#1''}%
\providecommand \bibnamefont  [1]{#1}%
\providecommand \bibfnamefont [1]{#1}%
\providecommand \citenamefont [1]{#1}%
\providecommand \href@noop [0]{\@secondoftwo}%
\providecommand \href [0]{\begingroup \@sanitize@url \@href}%
\providecommand \@href[1]{\@@startlink{#1}\@@href}%
\providecommand \@@href[1]{\endgroup#1\@@endlink}%
\providecommand \@sanitize@url [0]{\catcode `\\12\catcode `\$12\catcode
  `\&12\catcode `\#12\catcode `\^12\catcode `\_12\catcode `\%12\relax}%
\providecommand \@@startlink[1]{}%
\providecommand \@@endlink[0]{}%
\providecommand \url  [0]{\begingroup\@sanitize@url \@url }%
\providecommand \@url [1]{\endgroup\@href {#1}{\urlprefix }}%
\providecommand \urlprefix  [0]{URL }%
\providecommand \Eprint [0]{\href }%
\providecommand \doibase [0]{http://dx.doi.org/}%
\providecommand \selectlanguage [0]{\@gobble}%
\providecommand \bibinfo  [0]{\@secondoftwo}%
\providecommand \bibfield  [0]{\@secondoftwo}%
\providecommand \translation [1]{[#1]}%
\providecommand \BibitemOpen [0]{}%
\providecommand \bibitemStop [0]{}%
\providecommand \bibitemNoStop [0]{.\EOS\space}%
\providecommand \EOS [0]{\spacefactor3000\relax}%
\providecommand \BibitemShut  [1]{\csname bibitem#1\endcsname}%
\let\auto@bib@innerbib\@empty
\bibitem [{\citenamefont {Roca-Maza}\ \emph {et~al.}(2011)\citenamefont
  {Roca-Maza}, \citenamefont {Centelles}, \citenamefont {Vinas},\ and\
  \citenamefont {Warda}}]{RocaMaza:2011pm}%
  \BibitemOpen
  \bibfield  {author} {\bibinfo {author} {\bibfnamefont {X.}~\bibnamefont
  {Roca-Maza}}, \bibinfo {author} {\bibfnamefont {M.}~\bibnamefont
  {Centelles}}, \bibinfo {author} {\bibfnamefont {X.}~\bibnamefont {Vinas}}, \
  and\ \bibinfo {author} {\bibfnamefont {M.}~\bibnamefont {Warda}},\ }\href
  {\doibase 10.1103/PhysRevLett.106.252501} {\bibfield  {journal} {\bibinfo
  {journal} {Phys. Rev. Lett.}\ }\textbf {\bibinfo {volume} {106}},\ \bibinfo
  {pages} {252501} (\bibinfo {year} {2011})},\ \Eprint
  {http://arxiv.org/abs/1103.1762} {arXiv:1103.1762 [nucl-th]} \BibitemShut
  {NoStop}%
\bibitem [{\citenamefont {Roca-Maza}\ and\ \citenamefont
  {Paar}(2018)}]{Roca-Maza:2018ujj}%
  \BibitemOpen
  \bibfield  {author} {\bibinfo {author} {\bibfnamefont {X.}~\bibnamefont
  {Roca-Maza}}\ and\ \bibinfo {author} {\bibfnamefont {N.}~\bibnamefont
  {Paar}},\ }\href {\doibase 10.1016/j.ppnp.2018.04.001} {\bibfield  {journal}
  {\bibinfo  {journal} {Prog. Part. Nucl. Phys.}\ }\textbf {\bibinfo {volume}
  {101}},\ \bibinfo {pages} {96} (\bibinfo {year} {2018})},\ \Eprint
  {http://arxiv.org/abs/1804.06256} {arXiv:1804.06256 [nucl-th]} \BibitemShut
  {NoStop}%
\bibitem [{\citenamefont {Tsang}\ \emph {et~al.}(2012)\citenamefont {Tsang}
  \emph {et~al.}}]{Tsang:2012se}%
  \BibitemOpen
  \bibfield  {author} {\bibinfo {author} {\bibfnamefont {M.~B.}\ \bibnamefont
  {Tsang}} \emph {et~al.},\ }\href {\doibase 10.1103/PhysRevC.86.015803}
  {\bibfield  {journal} {\bibinfo  {journal} {Phys. Rev. C}\ }\textbf {\bibinfo
  {volume} {86}},\ \bibinfo {pages} {015803} (\bibinfo {year} {2012})},\
  \Eprint {http://arxiv.org/abs/1204.0466} {arXiv:1204.0466 [nucl-ex]}
  \BibitemShut {NoStop}%
\bibitem [{\citenamefont {Starodubsky}\ and\ \citenamefont
  {Hintz}(1994)}]{Starodubsky:1994xt}%
  \BibitemOpen
  \bibfield  {author} {\bibinfo {author} {\bibfnamefont {V.~E.}\ \bibnamefont
  {Starodubsky}}\ and\ \bibinfo {author} {\bibfnamefont {N.~M.}\ \bibnamefont
  {Hintz}},\ }\href {\doibase 10.1103/PhysRevC.49.2118} {\bibfield  {journal}
  {\bibinfo  {journal} {Phys. Rev. C}\ }\textbf {\bibinfo {volume} {49}},\
  \bibinfo {pages} {2118} (\bibinfo {year} {1994})}\BibitemShut {NoStop}%
\bibitem [{\citenamefont {Zenihiro}\ \emph {et~al.}(2010)\citenamefont
  {Zenihiro} \emph {et~al.}}]{Zenihiro:2010zz}%
  \BibitemOpen
  \bibfield  {author} {\bibinfo {author} {\bibfnamefont {J.}~\bibnamefont
  {Zenihiro}} \emph {et~al.},\ }\href {\doibase 10.1103/PhysRevC.82.044611}
  {\bibfield  {journal} {\bibinfo  {journal} {Phys. Rev. C}\ }\textbf {\bibinfo
  {volume} {82}},\ \bibinfo {pages} {044611} (\bibinfo {year}
  {2010})}\BibitemShut {NoStop}%
\bibitem [{\citenamefont {Klos}\ \emph {et~al.}(2007)\citenamefont {Klos} \emph
  {et~al.}}]{Klos:2007is}%
  \BibitemOpen
  \bibfield  {author} {\bibinfo {author} {\bibfnamefont {B.}~\bibnamefont
  {Klos}} \emph {et~al.},\ }\href {\doibase 10.1103/PhysRevC.76.014311}
  {\bibfield  {journal} {\bibinfo  {journal} {Phys. Rev. C}\ }\textbf {\bibinfo
  {volume} {76}},\ \bibinfo {pages} {014311} (\bibinfo {year} {2007})},\
  \Eprint {http://arxiv.org/abs/nucl-ex/0702016} {arXiv:nucl-ex/0702016}
  \BibitemShut {NoStop}%
\bibitem [{\citenamefont {Abrahamyan}\ \emph {et~al.}(2012)\citenamefont
  {Abrahamyan} \emph {et~al.}}]{Abrahamyan:2012gp}%
  \BibitemOpen
  \bibfield  {author} {\bibinfo {author} {\bibfnamefont {S.}~\bibnamefont
  {Abrahamyan}} \emph {et~al.},\ }\href {\doibase
  10.1103/PhysRevLett.108.112502} {\bibfield  {journal} {\bibinfo  {journal}
  {Phys. Rev. Lett.}\ }\textbf {\bibinfo {volume} {108}},\ \bibinfo {pages}
  {112502} (\bibinfo {year} {2012})},\ \Eprint {http://arxiv.org/abs/1201.2568}
  {arXiv:1201.2568 [nucl-ex]} \BibitemShut {NoStop}%
\bibitem [{\citenamefont {Friedman}(2012)}]{Friedman:2012pa}%
  \BibitemOpen
  \bibfield  {author} {\bibinfo {author} {\bibfnamefont {E.}~\bibnamefont
  {Friedman}},\ }\href {\doibase 10.1016/j.nuclphysa.2012.09.007} {\bibfield
  {journal} {\bibinfo  {journal} {Nucl. Phys. A}\ }\textbf {\bibinfo {volume}
  {896}},\ \bibinfo {pages} {46} (\bibinfo {year} {2012})},\ \Eprint
  {http://arxiv.org/abs/1209.6168} {arXiv:1209.6168 [nucl-ex]} \BibitemShut
  {NoStop}%
\bibitem [{\citenamefont {Tarbert}\ \emph {et~al.}(2014)\citenamefont {Tarbert}
  \emph {et~al.}}]{Tarbert:2013jze}%
  \BibitemOpen
  \bibfield  {author} {\bibinfo {author} {\bibfnamefont {C.~M.}\ \bibnamefont
  {Tarbert}} \emph {et~al.},\ }\href {\doibase 10.1103/PhysRevLett.112.242502}
  {\bibfield  {journal} {\bibinfo  {journal} {Phys. Rev. Lett.}\ }\textbf
  {\bibinfo {volume} {112}},\ \bibinfo {pages} {242502} (\bibinfo {year}
  {2014})},\ \Eprint {http://arxiv.org/abs/1311.0168} {arXiv:1311.0168
  [nucl-ex]} \BibitemShut {NoStop}%
\bibitem [{\citenamefont {Tamii}\ \emph {et~al.}(2011)\citenamefont {Tamii}
  \emph {et~al.}}]{Tamii:2011pv}%
  \BibitemOpen
  \bibfield  {author} {\bibinfo {author} {\bibfnamefont {A.}~\bibnamefont
  {Tamii}} \emph {et~al.},\ }\href {\doibase 10.1103/PhysRevLett.107.062502}
  {\bibfield  {journal} {\bibinfo  {journal} {Phys. Rev. Lett.}\ }\textbf
  {\bibinfo {volume} {107}},\ \bibinfo {pages} {062502} (\bibinfo {year}
  {2011})},\ \Eprint {http://arxiv.org/abs/1104.5431} {arXiv:1104.5431
  [nucl-ex]} \BibitemShut {NoStop}%
\bibitem [{\citenamefont {Piekarewicz}\ \emph {et~al.}(2012)\citenamefont
  {Piekarewicz}, \citenamefont {Agrawal}, \citenamefont {Colo}, \citenamefont
  {Nazarewicz}, \citenamefont {Paar}, \citenamefont {Reinhard}, \citenamefont
  {Roca-Maza},\ and\ \citenamefont {Vretenar}}]{Piekarewicz:2012pp}%
  \BibitemOpen
  \bibfield  {author} {\bibinfo {author} {\bibfnamefont {J.}~\bibnamefont
  {Piekarewicz}}, \bibinfo {author} {\bibfnamefont {B.~K.}\ \bibnamefont
  {Agrawal}}, \bibinfo {author} {\bibfnamefont {G.}~\bibnamefont {Colo}},
  \bibinfo {author} {\bibfnamefont {W.}~\bibnamefont {Nazarewicz}}, \bibinfo
  {author} {\bibfnamefont {N.}~\bibnamefont {Paar}}, \bibinfo {author}
  {\bibfnamefont {P.~G.}\ \bibnamefont {Reinhard}}, \bibinfo {author}
  {\bibfnamefont {X.}~\bibnamefont {Roca-Maza}}, \ and\ \bibinfo {author}
  {\bibfnamefont {D.}~\bibnamefont {Vretenar}},\ }\href {\doibase
  10.1103/PhysRevC.85.041302} {\bibfield  {journal} {\bibinfo  {journal} {Phys.
  Rev. C}\ }\textbf {\bibinfo {volume} {85}},\ \bibinfo {pages} {041302}
  (\bibinfo {year} {2012})},\ \Eprint {http://arxiv.org/abs/1201.3807}
  {arXiv:1201.3807 [nucl-th]} \BibitemShut {NoStop}%
\bibitem [{\citenamefont {Centelles}\ \emph {et~al.}(2010)\citenamefont
  {Centelles}, \citenamefont {Roca-Maza}, \citenamefont {Vinas},\ and\
  \citenamefont {Warda}}]{Centelles:2010qh}%
  \BibitemOpen
  \bibfield  {author} {\bibinfo {author} {\bibfnamefont {M.}~\bibnamefont
  {Centelles}}, \bibinfo {author} {\bibfnamefont {X.}~\bibnamefont
  {Roca-Maza}}, \bibinfo {author} {\bibfnamefont {X.}~\bibnamefont {Vinas}}, \
  and\ \bibinfo {author} {\bibfnamefont {M.}~\bibnamefont {Warda}},\ }\href
  {\doibase 10.1103/PhysRevC.82.054314} {\bibfield  {journal} {\bibinfo
  {journal} {Phys. Rev. C}\ }\textbf {\bibinfo {volume} {82}},\ \bibinfo
  {pages} {054314} (\bibinfo {year} {2010})},\ \Eprint
  {http://arxiv.org/abs/1010.5396} {arXiv:1010.5396 [nucl-th]} \BibitemShut
  {NoStop}%
\bibitem [{\citenamefont {Ray}\ \emph {et~al.}(1978)\citenamefont {Ray},
  \citenamefont {Coker},\ and\ \citenamefont {Hoffmann}}]{Ray:1978ws}%
  \BibitemOpen
  \bibfield  {author} {\bibinfo {author} {\bibfnamefont {L.}~\bibnamefont
  {Ray}}, \bibinfo {author} {\bibfnamefont {W.~R.}\ \bibnamefont {Coker}}, \
  and\ \bibinfo {author} {\bibfnamefont {G.~W.}\ \bibnamefont {Hoffmann}},\
  }\href {\doibase 10.1103/PhysRevC.18.2641} {\bibfield  {journal} {\bibinfo
  {journal} {Phys. Rev. C}\ }\textbf {\bibinfo {volume} {18}},\ \bibinfo
  {pages} {2641} (\bibinfo {year} {1978})}\BibitemShut {NoStop}%
\bibitem [{\citenamefont {Ray}(1979)}]{Ray:1979qv}%
  \BibitemOpen
  \bibfield  {author} {\bibinfo {author} {\bibfnamefont {L.}~\bibnamefont
  {Ray}},\ }\href {\doibase 10.1103/PhysRevC.20.1212} {\bibfield  {journal}
  {\bibinfo  {journal} {Phys. Rev. C}\ }\textbf {\bibinfo {volume} {19}},\
  \bibinfo {pages} {1855} (\bibinfo {year} {1979})},\ \bibinfo {note}
  {[Erratum: Phys.Rev.C 20, 1212--1212 (1979)]}\BibitemShut {NoStop}%
\bibitem [{\citenamefont {Hoffmann}\ \emph {et~al.}(1980)\citenamefont
  {Hoffmann} \emph {et~al.}}]{Hoffmann:1980kg}%
  \BibitemOpen
  \bibfield  {author} {\bibinfo {author} {\bibfnamefont {G.~W.}\ \bibnamefont
  {Hoffmann}} \emph {et~al.},\ }\href {\doibase 10.1103/PhysRevC.21.1488}
  {\bibfield  {journal} {\bibinfo  {journal} {Phys. Rev. C}\ }\textbf {\bibinfo
  {volume} {21}},\ \bibinfo {pages} {1488} (\bibinfo {year}
  {1980})}\BibitemShut {NoStop}%
\bibitem [{\citenamefont {Horowitz}(1985)}]{Horowitz:1985tw}%
  \BibitemOpen
  \bibfield  {author} {\bibinfo {author} {\bibfnamefont {C.~J.}\ \bibnamefont
  {Horowitz}},\ }\href {\doibase 10.1103/PhysRevC.31.1340} {\bibfield
  {journal} {\bibinfo  {journal} {Phys. Rev. C}\ }\textbf {\bibinfo {volume}
  {31}},\ \bibinfo {pages} {1340} (\bibinfo {year} {1985})}\BibitemShut
  {NoStop}%
\bibitem [{\citenamefont {Murdock}\ and\ \citenamefont
  {Horowitz}(1987)}]{Murdock:1986fs}%
  \BibitemOpen
  \bibfield  {author} {\bibinfo {author} {\bibfnamefont {D.~P.}\ \bibnamefont
  {Murdock}}\ and\ \bibinfo {author} {\bibfnamefont {C.~J.}\ \bibnamefont
  {Horowitz}},\ }\href {\doibase 10.1103/PhysRevC.35.1442} {\bibfield
  {journal} {\bibinfo  {journal} {Phys. Rev. C}\ }\textbf {\bibinfo {volume}
  {35}},\ \bibinfo {pages} {1442} (\bibinfo {year} {1987})}\BibitemShut
  {NoStop}%
\bibitem [{\citenamefont {Horowitz}\ \emph {et~al.}(1991)\citenamefont
  {Horowitz}, \citenamefont {Murdock},\ and\ \citenamefont
  {B.D.}}]{RIAcode:1991}%
  \BibitemOpen
  \bibfield  {author} {\bibinfo {author} {\bibfnamefont {C.}~\bibnamefont
  {Horowitz}}, \bibinfo {author} {\bibfnamefont {D.}~\bibnamefont {Murdock}}, \
  and\ \bibinfo {author} {\bibfnamefont {S.}~\bibnamefont {B.D.}},\ }\href@noop
  {} {\emph {\bibinfo {title} {Computational Nuclear Physics 1}}},\ edited by\
  \bibinfo {editor} {\bibfnamefont {K.}~\bibnamefont {Langanke}}, \bibinfo
  {editor} {\bibfnamefont {J.}~\bibnamefont {Maruhn}}, \ and\ \bibinfo {editor}
  {\bibfnamefont {S.}~\bibnamefont {Koonin}}\ (\bibinfo  {publisher}
  {Springer-Verlag},\ \bibinfo {year} {1991})\ p.\ \bibinfo {pages}
  {129}\BibitemShut {NoStop}%
\bibitem [{\citenamefont {Sakaguchi}\ \emph {et~al.}(1998)\citenamefont
  {Sakaguchi} \emph {et~al.}}]{Sakaguchi:1998zz}%
  \BibitemOpen
  \bibfield  {author} {\bibinfo {author} {\bibfnamefont {H.}~\bibnamefont
  {Sakaguchi}} \emph {et~al.},\ }\href {\doibase 10.1103/PhysRevC.57.1749}
  {\bibfield  {journal} {\bibinfo  {journal} {Phys. Rev. C}\ }\textbf {\bibinfo
  {volume} {57}},\ \bibinfo {pages} {1749} (\bibinfo {year}
  {1998})}\BibitemShut {NoStop}%
\bibitem [{\citenamefont {Terashima}\ \emph {et~al.}(2008)\citenamefont
  {Terashima} \emph {et~al.}}]{Terashima:2008zza}%
  \BibitemOpen
  \bibfield  {author} {\bibinfo {author} {\bibfnamefont {S.}~\bibnamefont
  {Terashima}} \emph {et~al.},\ }\href {\doibase 10.1103/PhysRevC.77.024317}
  {\bibfield  {journal} {\bibinfo  {journal} {Phys. Rev. C}\ }\textbf {\bibinfo
  {volume} {77}},\ \bibinfo {pages} {024317} (\bibinfo {year}
  {2008})}\BibitemShut {NoStop}%
\bibitem [{\citenamefont {Zenihiro}\ \emph {et~al.}(2018)\citenamefont
  {Zenihiro} \emph {et~al.}}]{Zenihiro:2018rmz}%
  \BibitemOpen
  \bibfield  {author} {\bibinfo {author} {\bibfnamefont {J.}~\bibnamefont
  {Zenihiro}} \emph {et~al.},\ }\href@noop {} {\  (\bibinfo {year} {2018})},\
  \Eprint {http://arxiv.org/abs/1810.11796} {arXiv:1810.11796 [nucl-ex]}
  \BibitemShut {NoStop}%
\bibitem [{\citenamefont {Piekarewicz}\ and\ \citenamefont
  {Weppner}(2006)}]{Piekarewicz:2005iu}%
  \BibitemOpen
  \bibfield  {author} {\bibinfo {author} {\bibfnamefont {J.}~\bibnamefont
  {Piekarewicz}}\ and\ \bibinfo {author} {\bibfnamefont {S.~P.}\ \bibnamefont
  {Weppner}},\ }\href {\doibase 10.1016/j.nuclphysa.2006.08.004} {\bibfield
  {journal} {\bibinfo  {journal} {Nucl. Phys. A}\ }\textbf {\bibinfo {volume}
  {778}},\ \bibinfo {pages} {10} (\bibinfo {year} {2006})},\ \Eprint
  {http://arxiv.org/abs/nucl-th/0509019} {arXiv:nucl-th/0509019} \BibitemShut
  {NoStop}%
\bibitem [{\citenamefont {Yoshida}\ \emph {et~al.}(2020)\citenamefont
  {Yoshida}, \citenamefont {Sagawa}, \citenamefont {Zenihiro},\ and\
  \citenamefont {Uesaka}}]{PhysRevC.102.064307}%
  \BibitemOpen
  \bibfield  {author} {\bibinfo {author} {\bibfnamefont {S.}~\bibnamefont
  {Yoshida}}, \bibinfo {author} {\bibfnamefont {H.}~\bibnamefont {Sagawa}},
  \bibinfo {author} {\bibfnamefont {J.}~\bibnamefont {Zenihiro}}, \ and\
  \bibinfo {author} {\bibfnamefont {T.}~\bibnamefont {Uesaka}},\ }\href
  {\doibase 10.1103/PhysRevC.102.064307} {\bibfield  {journal} {\bibinfo
  {journal} {Phys. Rev. C}\ }\textbf {\bibinfo {volume} {102}},\ \bibinfo
  {pages} {064307} (\bibinfo {year} {2020})}\BibitemShut {NoStop}%
\bibitem [{\citenamefont {Roca-Maza}\ \emph {et~al.}(2008)\citenamefont
  {Roca-Maza}, \citenamefont {Centelles}, \citenamefont {Salvat},\ and\
  \citenamefont {Vinas}}]{RocaMaza:2008cg}%
  \BibitemOpen
  \bibfield  {author} {\bibinfo {author} {\bibfnamefont {X.}~\bibnamefont
  {Roca-Maza}}, \bibinfo {author} {\bibfnamefont {M.}~\bibnamefont
  {Centelles}}, \bibinfo {author} {\bibfnamefont {F.}~\bibnamefont {Salvat}}, \
  and\ \bibinfo {author} {\bibfnamefont {X.}~\bibnamefont {Vinas}},\ }\href
  {\doibase 10.1103/PhysRevC.78.044332} {\bibfield  {journal} {\bibinfo
  {journal} {Phys. Rev. C}\ }\textbf {\bibinfo {volume} {78}},\ \bibinfo
  {pages} {044332} (\bibinfo {year} {2008})},\ \Eprint
  {http://arxiv.org/abs/0808.1252} {arXiv:0808.1252 [nucl-th]} \BibitemShut
  {NoStop}%
\bibitem [{\citenamefont {Hoffmann}\ \emph {et~al.}(1990)\citenamefont
  {Hoffmann} \emph {et~al.}}]{Hoffmann:1990ve}%
  \BibitemOpen
  \bibfield  {author} {\bibinfo {author} {\bibfnamefont {G.~W.}\ \bibnamefont
  {Hoffmann}} \emph {et~al.},\ }\href {\doibase 10.1103/PhysRevC.41.1651}
  {\bibfield  {journal} {\bibinfo  {journal} {Phys. Rev. C}\ }\textbf {\bibinfo
  {volume} {41}},\ \bibinfo {pages} {1651} (\bibinfo {year}
  {1990})}\BibitemShut {NoStop}%
\bibitem [{\citenamefont {Nik\v{s}i\'{c}}\ \emph {et~al.}(2014)\citenamefont
  {Nik\v{s}i\'{c}}, \citenamefont {Paar}, \citenamefont {Vretenar},\ and\
  \citenamefont {Ring}}]{Niksic:2014dra}%
  \BibitemOpen
  \bibfield  {author} {\bibinfo {author} {\bibfnamefont {T.}~\bibnamefont
  {Nik\v{s}i\'{c}}}, \bibinfo {author} {\bibfnamefont {N.}~\bibnamefont
  {Paar}}, \bibinfo {author} {\bibfnamefont {D.}~\bibnamefont {Vretenar}}, \
  and\ \bibinfo {author} {\bibfnamefont {P.}~\bibnamefont {Ring}},\ }\href
  {\doibase 10.1016/j.cpc.2014.02.027} {\bibfield  {journal} {\bibinfo
  {journal} {Comput. Phys. Commun.}\ }\textbf {\bibinfo {volume} {185}},\
  \bibinfo {pages} {1808} (\bibinfo {year} {2014})},\ \Eprint
  {http://arxiv.org/abs/1403.4039} {arXiv:1403.4039 [nucl-th]} \BibitemShut
  {NoStop}%
\bibitem [{\citenamefont {Lalazissis}\ \emph {et~al.}(2005)\citenamefont
  {Lalazissis}, \citenamefont {Nik\v{s}i\'{c}}, \citenamefont {Vretenar},\ and\
  \citenamefont {Ring}}]{Lalazissis:2005de}%
  \BibitemOpen
  \bibfield  {author} {\bibinfo {author} {\bibfnamefont {G.~A.}\ \bibnamefont
  {Lalazissis}}, \bibinfo {author} {\bibfnamefont {T.}~\bibnamefont
  {Nik\v{s}i\'{c}}}, \bibinfo {author} {\bibfnamefont {D.}~\bibnamefont
  {Vretenar}}, \ and\ \bibinfo {author} {\bibfnamefont {P.}~\bibnamefont
  {Ring}},\ }\href {\doibase 10.1103/PhysRevC.71.024312} {\bibfield  {journal}
  {\bibinfo  {journal} {Phys. Rev. C}\ }\textbf {\bibinfo {volume} {71}},\
  \bibinfo {pages} {024312} (\bibinfo {year} {2005})}\BibitemShut {NoStop}%
\bibitem [{\citenamefont {Nik\v{s}i\'{c}}\ \emph {et~al.}(2008)\citenamefont
  {Nik\v{s}i\'{c}}, \citenamefont {Vretenar},\ and\ \citenamefont
  {Ring}}]{Niksic:2008vp}%
  \BibitemOpen
  \bibfield  {author} {\bibinfo {author} {\bibfnamefont {T.}~\bibnamefont
  {Nik\v{s}i\'{c}}}, \bibinfo {author} {\bibfnamefont {D.}~\bibnamefont
  {Vretenar}}, \ and\ \bibinfo {author} {\bibfnamefont {P.}~\bibnamefont
  {Ring}},\ }\href {\doibase 10.1103/PhysRevC.78.034318} {\bibfield  {journal}
  {\bibinfo  {journal} {Phys. Rev. C}\ }\textbf {\bibinfo {volume} {78}},\
  \bibinfo {pages} {034318} (\bibinfo {year} {2008})},\ \Eprint
  {http://arxiv.org/abs/0809.1375} {arXiv:0809.1375 [nucl-th]} \BibitemShut
  {NoStop}%
\bibitem [{\citenamefont {Bennaceur}\ and\ \citenamefont
  {Dobaczewski}(2005)}]{Bennaceur:2005mx}%
  \BibitemOpen
  \bibfield  {author} {\bibinfo {author} {\bibfnamefont {K.}~\bibnamefont
  {Bennaceur}}\ and\ \bibinfo {author} {\bibfnamefont {J.}~\bibnamefont
  {Dobaczewski}},\ }\href {\doibase 10.1016/j.cpc.2005.02.002} {\bibfield
  {journal} {\bibinfo  {journal} {Comput. Phys. Commun.}\ }\textbf {\bibinfo
  {volume} {168}},\ \bibinfo {pages} {96} (\bibinfo {year} {2005})},\ \Eprint
  {http://arxiv.org/abs/nucl-th/0501002} {arXiv:nucl-th/0501002} \BibitemShut
  {NoStop}%
\bibitem [{\citenamefont {Bartel}\ \emph {et~al.}(1982)\citenamefont {Bartel},
  \citenamefont {Quentin}, \citenamefont {Brack}, \citenamefont {Guet},\ and\
  \citenamefont {Hakansson}}]{Bartel:1982ed}%
  \BibitemOpen
  \bibfield  {author} {\bibinfo {author} {\bibfnamefont {J.}~\bibnamefont
  {Bartel}}, \bibinfo {author} {\bibfnamefont {P.}~\bibnamefont {Quentin}},
  \bibinfo {author} {\bibfnamefont {M.}~\bibnamefont {Brack}}, \bibinfo
  {author} {\bibfnamefont {C.}~\bibnamefont {Guet}}, \ and\ \bibinfo {author}
  {\bibfnamefont {H.~B.}\ \bibnamefont {Hakansson}},\ }\href {\doibase
  10.1016/0375-9474(82)90403-1} {\bibfield  {journal} {\bibinfo  {journal}
  {Nucl. Phys. A}\ }\textbf {\bibinfo {volume} {386}},\ \bibinfo {pages} {79}
  (\bibinfo {year} {1982})}\BibitemShut {NoStop}%
\bibitem [{\citenamefont {Chabanat}\ \emph {et~al.}(1998)\citenamefont
  {Chabanat}, \citenamefont {Bonche}, \citenamefont {Haensel}, \citenamefont
  {Meyer},\ and\ \citenamefont {Schaeffer}}]{Chabanat:1997un}%
  \BibitemOpen
  \bibfield  {author} {\bibinfo {author} {\bibfnamefont {E.}~\bibnamefont
  {Chabanat}}, \bibinfo {author} {\bibfnamefont {P.}~\bibnamefont {Bonche}},
  \bibinfo {author} {\bibfnamefont {P.}~\bibnamefont {Haensel}}, \bibinfo
  {author} {\bibfnamefont {J.}~\bibnamefont {Meyer}}, \ and\ \bibinfo {author}
  {\bibfnamefont {R.}~\bibnamefont {Schaeffer}},\ }\href {\doibase
  10.1016/S0375-9474(98)00180-8} {\bibfield  {journal} {\bibinfo  {journal}
  {Nucl. Phys. A}\ }\textbf {\bibinfo {volume} {635}},\ \bibinfo {pages} {231}
  (\bibinfo {year} {1998})},\ \bibinfo {note} {[Erratum: Nucl.Phys.A 643,
  441--441 (1998)]}\BibitemShut {NoStop}%
\bibitem [{\citenamefont {Zenihiro}(2011)}]{Zenihiro:dron}%
  \BibitemOpen
  \bibfield  {author} {\bibinfo {author} {\bibfnamefont {J.}~\bibnamefont
  {Zenihiro}},\ }\emph {\bibinfo {title} {Neutron density distributions of Pb
  deduced via proton elastic scattering at 295 MeV}},\ \href@noop {} {Ph.D.
  thesis},\ \bibinfo  {school} {Kyoto University} (\bibinfo {year}
  {2011})\BibitemShut {NoStop}%
\bibitem [{\citenamefont {De~Vries}\ \emph {et~al.}(1987)\citenamefont
  {De~Vries}, \citenamefont {De~Jager},\ and\ \citenamefont
  {De~Vries}}]{DeJager:1987qc}%
  \BibitemOpen
  \bibfield  {author} {\bibinfo {author} {\bibfnamefont {H.}~\bibnamefont
  {De~Vries}}, \bibinfo {author} {\bibfnamefont {C.~W.}\ \bibnamefont
  {De~Jager}}, \ and\ \bibinfo {author} {\bibfnamefont {C.}~\bibnamefont
  {De~Vries}},\ }\href {\doibase 10.1016/0092-640X(87)90013-1} {\bibfield
  {journal} {\bibinfo  {journal} {Atom. Data Nucl. Data Tabl.}\ }\textbf
  {\bibinfo {volume} {36}},\ \bibinfo {pages} {495} (\bibinfo {year}
  {1987})}\BibitemShut {NoStop}%
\end{thebibliography}%

\end{document}